\documentclass[preprint,12pt]{aastex}
\usepackage{latexsym}
\usepackage{amsmath}
\usepackage{amsfonts}
\usepackage{amssymb}
\usepackage{makeidx}
\usepackage{graphicx}
\usepackage{natbib}
\usepackage{footnote}
\usepackage[margin=0.7in]{geometry}

\newcommand{\um}{\mbox{\,$\mu$m}}
\newcommand{\nh}{N$_{2}$H$^{+}$(1-0)}
\newcommand{\hcop}{HCO$^{\text{+}}$(1-0)}
\newcommand{\hcn}{HCN(1-0)}

\begin{document}
\title{CARMA Large Area Star Formation Survey: Structure and Kinematics of Dense Gas in Serpens Main}
\author{Katherine I.\ Lee\altaffilmark{1,2}, 
Manuel Fern\'{a}ndez-L\'{o}pez\altaffilmark{2,3}, 
Shaye Storm\altaffilmark{1}, 
Leslie W.\ Looney\altaffilmark{2}, 
Lee G.\ Mundy\altaffilmark{1},
Dominique Segura-Cox\altaffilmark{2},
Peter Teuben\altaffilmark{1},
Erik Rosolowsky\altaffilmark{4,5},
H\'{e}ctor G.\ Arce\altaffilmark{6},
Eve C.\ Ostriker\altaffilmark{7},
Yancy L.\ Shirley\altaffilmark{8},
Woojin Kwon\altaffilmark{9},
Jens Kauffmann\altaffilmark{10},
John J.\ Tobin\altaffilmark{11},
Adele L.\ Plunkett\altaffilmark{6},
Marc W.\ Pound\altaffilmark{1},
Demerese M.\ Salter\altaffilmark{1},
N.\ H.\ Volgenau\altaffilmark{12,15},
Che-Yu Chen\altaffilmark{1},
Konstantinos Tassis\altaffilmark{13,14},
Andrea Isella\altaffilmark{12},
Richard M.\ Crutcher\altaffilmark{2},
Charles F.\ Gammie\altaffilmark{2},
Leonardo Testi\altaffilmark{16}
}

\altaffiltext{1}{Department of Astronomy, University of Maryland, College Park, MD 20742, USA; ijlee9@astro.umd.edu}
\altaffiltext{2}{Department of Astronomy, University of Illinois, Urbana-Champaign, IL 61801, USA}
\altaffiltext{3}{Instituto Argentino de Radioastronom\'{i}a, CCT-La Plata (CONICET), C.C.5, 1894, Villa Elisa, Argentina}
\altaffiltext{4}{University of British Columbia, Okanagan Campus, Departments of Physics and Statistics, 3333 University Way, Kelowna BC V1V 1V7, Canada}
\altaffiltext{5}{University of Alberta, Department of Physics, 4-181 CCIS, Edmonton AB T6G 2E1, Canada}
\altaffiltext{6}{Department of Astronomy, Yale University, P.O. Box 208101, New Haven, CT 06520-8101, USA}
\altaffiltext{7}{Department of Astrophysical Sciences, Princeton University, Princeton, NJ 08544, USA}
\altaffiltext{8}{Steward Observatory, 933 North Cherry Avenue, Tucson, AZ 85721, USA}
\altaffiltext{9}{SRON Netherlands Institute for Space Research, Landleven 12, 9747 AD Groningen, The Netherlands}
\altaffiltext{10}{Max Planck Institut f$\ddot{\text{u}}$r Radioastronomie, Auf dem H$\ddot{\text{u}}$gel 69 D53121, Bonn Germany}
\altaffiltext{11}{National Radio Astronomy Observatory, Charlottesville, VA 22903, USA}
\altaffiltext{12}{Astronomy Department, California Institute of Technology, 1200 East California Blvd., Pasadena, CA 91125, USA}
\altaffiltext{13}{Department of Physics and Institute of Theoretical \& Computational Physics, University of Crete, PO Box 2208, GR-710 03, Heraklion, Crete, Greece}
\altaffiltext{14}{Foundation for Research and Technology - Hellas, IESL, Voutes, 7110 Heraklion, Greece}
\altaffiltext{15}{Owens Valley Radio Observatory, MC 105-24 OVRO, Pasadena, CA 91125, USA}
\altaffiltext{16}{ESO, Karl-Schwarzschild-Strasse 2 D-85748 Garching bei M$\ddot{\text{u}}$nchen, Germany}

\begin{abstract}

We present observations of N$_{2}$H$^{+}$ ($J=1 \rightarrow 0$), HCO$^{+}$ ($J=1 \rightarrow 0$), and HCN ($J=1 \rightarrow 0$) toward the Serpens Main molecular cloud from the CARMA Large Area Star Formation Survey (CLASSy). 
We mapped 150 square arcminutes of Serpens Main with an angular resolution of $\sim$ 7\arcsec.  
The gas emission is concentrated in two subclusters (the NW and SE subclusters). 
The SE subcluster has more prominent filamentary structures and more complicated kinematics compared to the NW subcluster. 
The majority of gas in the two subclusters has subsonic to sonic velocity dispersions. 
We applied a dendrogram technique with \nh \ to study the gas structures; 
the SE subcluster has a higher degree of hierarchy than the NW subcluster. 
Combining the dendrogram and line fitting analyses reveals two distinct relations:
a flat relation between nonthermal velocity dispersion and size, and a positive correlation between variation in velocity centroids and size.
The two relations imply a characteristic depth of 0.15~pc for the cloud. 
Furthermore, we have identified six filaments in the SE subcluster. 
These filaments have lengths of $\sim$~0.2 pc and widths of $\sim 0.03$~pc, which is smaller than a characteristic width of 0.1~pc suggested by \textit{Herschel} observations. 
The filaments can be classified into two types based on their properties. 
The first type, located in the northeast of the SE subcluster, has larger velocity gradients, smaller masses, and nearly critical mass-per-unit-length ratios. 
The other type, located in the southwest of the SE subcluster, has the opposite properties. 
Several YSOs are formed along two filaments which have supercritical mass per unit length ratios, while filaments with nearly critical mass-per-unit-length ratios are not associated with YSOs, 
suggesting that stars are formed on gravitationally unstable filaments.

\end{abstract}

\section{Introduction}

\subsection{The CLASSy Project}

The star formation process spans a wide range of spatial scales: 
from molecular clouds on  parsec scales, to envelopes around young stellar objects on few thousand AU scales, to circumstellar disks on 1 to 100 AU scales.
Low-density gas ($\sim 10^{2}$ cm$^{-3}$) in the interstellar medium forms denser structures in molecular clouds ($\sim 10^{3}$ cm$^{-3}$), which evolve to higher density structures ($\sim 10^{5}$ cm$^{-3}$) at small scales to enable the formation of stars and clusters.
This general picture represents the broad path from low density gas to star formation; however,
a comprehensive understanding of the roles of turbulence, magnetic fields, and gravity at all spatial scales in driving this evolution is still needed.
While several theoretical scenarios have been proposed to address this need \citep{1976ApJ...210..326M,1995ApJ...453..271B,2004RvMP...76..125M,2007ARA&A..45..565M},
observational tests have been insufficient to directly probe the conditions for star formation from few thousand AUs to parsecs to produce an integrated picture.
Previous surveys of nearby star forming regions have been carried out with the \textit{Herschel} Gould Belt Survey \citep[e.g.,][]{2010A&A...518L.102A}, the \textit{Spitzer} Legacy c2d project \citep[e.g.,][]{2003PASP..115..965E,2006ApJ...644..307H,2009ApJS..181..321E}, and the \textit{JCMT} Legacy Survey \citep[e.g.,][]{2010MNRAS.401..204B}.
These surveys have provided important insights into star formation from infrared to submillimeter regimes.
However, there has been a lack of large-area mapping (at parsec-scales) of the molecular gas with high angular resolution (few thousands of AU scales) and sensitivity \citep[c.f.,][]{2013ApJ...764L..26B} to probe the gas structure and density in detail.

The CARMA Large Area Star Formation Survey (CLASSy), a survey toward five star-forming regions in the nearby Gould Belt,  addresses this gap.
By combining the interferometric and single-dish data with the full CARMA 23 antennas,
CLASSy observed the emission from three high density gas tracer molecules in target regions over a broad range of spatial scales (from few parsecs to few thousand AUs): 
\nh, \hcop, and \hcn.
The primary goals of the survey are to: (1) characterize the internal structure and dynamics of star-forming cores, (2) investigate the relationship between dense cores and their natal molecular clouds, and (3) test theoretical scenarios for star formation.
The target regions, NGC 1333, Barnard 1, and L1451 in Perseus, and Serpens Main and Serpens South,
present a wide range of star formation activities from relatively quiescent regions to active star-forming clusters.
\citet{storm14} (hereafter Paper I) presents a detailed description on the CLASSy project and the structures of dense gas in Barnard 1.  
In this paper, we present results of the Serpens Main region.
We focus on the global structure of dust and gas, including the properties of dust and gas condensations, gas structures and kinematics, and filamentary structures.

\subsection{Serpens Main}

\begin{figure}[h]
\begin{center}
\includegraphics[scale=1.5]{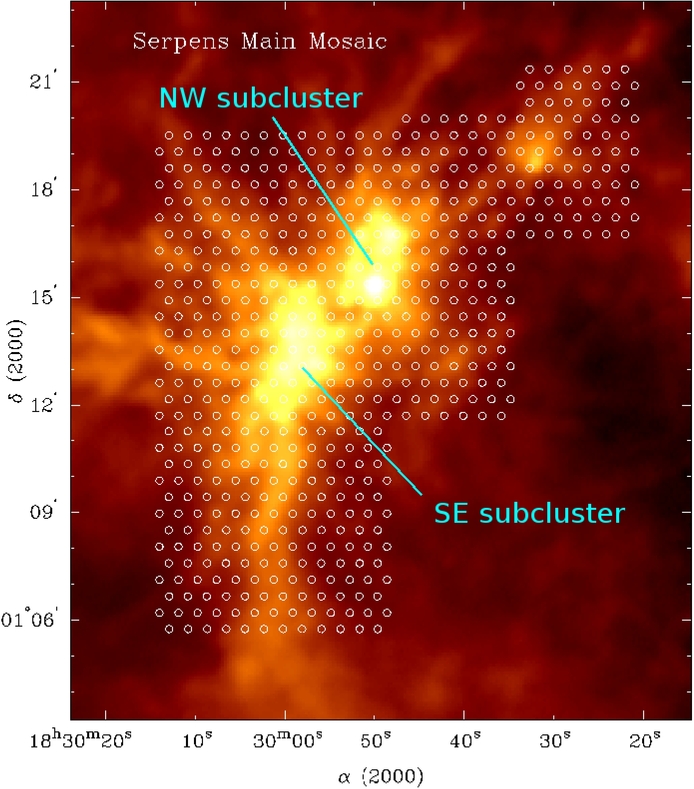}
\caption{\footnotesize The mosaic pointing centers (white circles) for the Serpens Main observations.  The color image is from \textit{Herschel} 250 \micron \ \citep{2010A&A...518L.102A}.  The primary beam at 90 GHz from the 10-m antennas is 77\arcsec.}
\label{fig:pointing}
\end{center}
\end{figure}

Serpens Main is a young cluster active in star formation \citep[see][and references therein]{2008hsf2.book..693E}.
The dust and gas in Serpens Main are mainly concentrated in two subclusters,     
the NW and SE subclusters \citep{1979ApJ...234..932L,1993A&A...275..195C,1999MNRAS.309..141D,2000ApJ...536..845M,2004A&A...421..623K}. 
Figure~\ref{fig:pointing} shows Serpens Main and the two subclusters at 250 \micron \ from \textit{Herschel} observations \citep{2010A&A...518L.102A}.
The gas mass estimated in the Serpens Main region is about $97$ M$_{\sun}$ in the NW subcluster and about 144 M$_{\sun}$ in the SE subcluster using a distance of 415~pc \citep{2002A&A...392.1053O}.

A few hundred YSOs were identified in Serpens Main based on infrared observations including the \textit{Spitzer} IRAC and MIPS bands \citep{2006ApJ...644..307H,2009ApJS..181..321E}, and ISOCAM \citep{2004A&A...421..623K}.
Previous studies have shown a high fraction of protostars to stars with disks in Serpens Main \citep{2004A&A...421..623K}. 
For example, \citet{2007ApJ...669..493W} discovered a high ratio of 48\% between protostars (22 Class 0/I sources, 16 flat-spectrum sources) and pre-main-sequence stars with disks (62 Class II sources, 17 transition disks) in Serpens Main. 
The younger sources (Class 0/I and flat-spectrum sources) are mostly found in the central NW and SE subclusters, 
while more evolved sources (Class II/III) are dispersed over the larger region \citep[e.g.,][]{2007ApJ...663.1139H}.
The star formation activities have been suggested to have undergone multiple phases \citep{2004A&A...421..623K,1993A&A...275..195C}: 
Class II and Class III sources were formed and dispersed $2\times 10^{6}$ yr ago, and the current burst of Class 0 and I sources started $\sim 10^{5}$ yr ago.
The SE subcluster is suggested to be more evolved than the NW subcluster due to its higher fraction of Class II/III YSOs \citep{2007ApJ...669..493W}.

Submillimeter to millimeter-wavelength observations have revealed a number of embedded sources as well.
\citet{1999MNRAS.309..141D} observed the Serpens Main region at 450 $\um$ and 850 $\um$ with the JCMT SCUBA array and identified eleven submillimeter sources. 
\citet{2010ApJ...710.1247S} found four starless cores in the field.
\citet{2007ApJ...666..982E} identified twelve continuum sources at 1~mm with Bolocam.   
Interferometric observations with high angular resolutions at 3~mm \citep{1998ApJ...508L..91T,2000ApJ...537..891W} found continuum sources coincident with the submillimeter cores.
These sources are concentrated in the two subclusters, 
and infall motions have been suggested toward several of these sources \citep{1997ApJ...484..256G,2000ApJ...537..891W,2002A&A...392.1053O}.
A number of outflows have been observed to be associated with these young sources \citep{1992PASJ...44..155E,1995A&A...298..594W,1998ApJ...501L.193W,1999MNRAS.309..141D,2000ApJ...540L..53T,2010MNRAS.409.1412G}. 

Despite some similarities in the dust emission in the two subclusters, 
the gas kinematics and temperatures are distinct \citep{2000ApJ...540L..53T,2010A&A...519A..27D}. 
The NW subcluster has a more uniform velocity field, 
while the SE subcluster has a large-scale velocity gradient in the E-W direction \citep{2010MNRAS.409.1412G}.
The velocity gradient was interpreted as rotation \citep{2002A&A...392.1053O},
but it has also been suggested to be the result of a cloud-cloud collision  \citep{2011A&A...528A..50D}.

There are various estimates in the literature for the gas temperature in Serpens Main.
Gas temperature in the two subclusters has been estimated between 12 K and 19 K based on NH$_{3}$ \citep{1984A&A...131..177U,1996ApJ...456..677C,2013A&A...553A..58L}.
Earlier works reported 25-27 K based on CO observations \citep{1979ApJ...234..932L}, and 22-35 K based on dust emission at infrared \citep{1994ApJ...424..222M,1996ApJ...456..686H}.
Based on C$^{17}$O(1-0) and C$^{18}$O(1-0) observations, \citep{2010A&A...519A..27D} concluded that the NW subcluster presents a homogeneous 10 K gas temperature, 
while the gas temperature in the SE subcluster is higher between 10 K and 20 K.
In this paper, we assume a mean gas temperature of 20 K for the overall cloud \citep{2000ApJ...536..845M}, 
and 13 K for regions that do not show active star formation activities \citep{2013A&A...553A..58L}. 

The distance to the Serpens region has been under debate \citep[see the discussion in][]{2008hsf2.book..693E,2010AJ....140..266W}. 
Several studies adopted $260 \pm 10$ pc from \citet{1996BaltA...5..125S} based on the spectral and luminosity class of the observed stars. 
Recently, a new estimate of $415 \pm 5$~pc was found based on trigonometric parallax with VLBI observations with the young stellar object EC 95~in the SE subcluster of Serpens Main \citep{2010ApJ...718..610D}. 
Given the accuracy of the VLBI observations, we adopt a distance of 415 pc in this paper.

\section{Observations}

\begin{deluxetable}{cccccccc}
\tablewidth{0pc}
\tablecolumns{8}
\tabletypesize{\tiny}
\tablecaption{Summary of Observations}
\tablehead{
\colhead{Array} & \colhead{Tracks} & \colhead{Hours} & \colhead{Flux} & \colhead{Gain}& \colhead{Baseline} & \colhead{Flux on gaincal} & \colhead{Dates} \\
\colhead{configuration} & \colhead{observed} & \colhead{observed} & \colhead{calibrators} & \colhead{calibrators}& \colhead{(k$\lambda$)} & \colhead{(Jy)}  & \colhead{} \\
}
\startdata
DZ & 10 & 54 & Neptune \& Uranus & J1743-038 & 1.1-45 & 3.8-4.3 & April - June 2012 \\
EZ & 10 & 47 & Neptune & J1743-038 & 1.1-30 & 3.9-4.3 & July - August 2012 \\
\enddata
\label{tobs}
\end{deluxetable}

CARMA is a heterogeneous array that comprises 23 antennas (six 10.4-m, nine 6.1-m and eight 3.5-m dishes). 
Our observations used CARMA23 mode, which correlates all 23 antennas, providing 253 baselines ranging from 1.1 to 45 k$\lambda$ in DZ and EZ configurations.
The observations of Serpens Main were performed in the DZ and EZ configurations between April and August, 2012 (see Table \ref{tobs} for details).
We acquired 10 tracks under good atmospheric conditions in both DZ and EZ configurations, respectively, totaling an on-source integration time of $\sim 100$ hours. 
The region was mapped with a 531-pointing Nyquist-sampled mosaic with 31\arcsec \ spacing, covering about 150 square minutes. 
The pointing centers are shown in Figure \ref{fig:pointing}. 
We acquired total power observations to recover resolved-out line emission by observing an emission-free position (at $\alpha_{J2000}=18^h 31^m 02\fs2$, $\delta_{J2000}=+01\degr 23\arcmin 30\farcs0$) every 3.5 minutes for tracks with stable atmospheric opacity.

The CARMA23 correlator has four spectral bands in the upper side band. 
We configured one 500 MHz band with spectral resolution of 31.25 MHz for continuum observations, 
and three 8 MHz bands with spectral resolution of 0.050 MHz for molecular line observations (see Table 2 for details). 
The three molecular lines were placed in the center of each 8 MHz band, 
which provided velocity resolution $\sim$0.16~km/s and velocity coverage $\sim$24~km/s.

\begin{deluxetable}{cccccccc}
\tablewidth{0pc}
\tablecolumns{8}
\tabletypesize{\tiny}
\tablecaption{Correlator Setup Summary}
\tablehead{
\colhead{Line} & \colhead{Rest Freq.} & \colhead{No.\ Chan.} & \colhead{Chan.\ Width} & \colhead{Vel.\ Coverage}& \colhead{Vel.\ Resolution} & \colhead{Chan.\ RMS} & \colhead{Synth.\ Beam} \\
\colhead{} & \colhead{(GHz)} & \colhead{} & \colhead{(MHz)} & \colhead{(km~s$^{-1}$)}& \colhead{(km~s$^{-1}$)} & \colhead{(Jy~beam$^{-1}$)}  & \colhead{} \\
}
\startdata
\nh & 93.173704 & 159 & 0.049 & 24.82 & 0.157 & 0.20 & $7.7\arcsec \times 7.0\arcsec$ \\
Continuum & 92.7947 & 47 & 10.4 & 1547 & 33.6 & 0.0015 & $7.5\arcsec \times 7.0\arcsec$ \\
\hcop & 89.188518 & 159 & 0.049 & 25.92 & 0.164 & 0.18 & $7.2\arcsec \times 7.9\arcsec$ \\
\hcn & 88.631847 & 159 & 0.049 & 26.10 & 0.165 & 0.17 & $7.3\arcsec \times 8.0\arcsec$
\enddata
\label{tbl:corr}
\end{deluxetable}

The gain calibrator was the quasar J1743-038, which was observed every 20 minutes. 
Neptune was used as the primary flux standard, although Uranus was used in the first track. 
The upper limit for absolute flux calibration is 15\%.  

A detailed procedure of our data reduction is described in Paper I, and summarized here for Serpens Main.
The data were calibrated using the MIRIAD software package \citep{1995ASPC...77..433S}. 
The 500 MHz bandpass was calibrated using astronomical sources (3C279 and Neptune, which were unresolved in D and E configurations). 
For the 8 MHz bands, the CARMA built-in correlated noise source was used to calculate bandpass corrections \citep[e.g.,][]{2008SPIE.7018E..25W}.  
Interferometric data from the 10.4-m and 3.5-m baselines were removed because the first negative sidelobe of the 10.4-m beam illuminates the 3.5-m beam. 
Single-dish data from the 10.4-m antennas were calibrated and imaged using SINBAD, SINPOLY and VARMAPS in MIRIAD. 
For molecular line imaging, the interferometric and single-dish data were jointly deconvolved using MOSMEM to create the calibrated position-position-velocity cubes.
Only interferometric data was used in the continuum imaging since CARMA is not capable of the fast chopping required for single-dish continuum calibration.
Table~\ref{tbl:corr} summarizes the correlator setup, RMS levels and synthesized beams for the calibrated images.

\section{Results}

\begin{figure}[h]
\begin{center}
\includegraphics[scale=2.4]{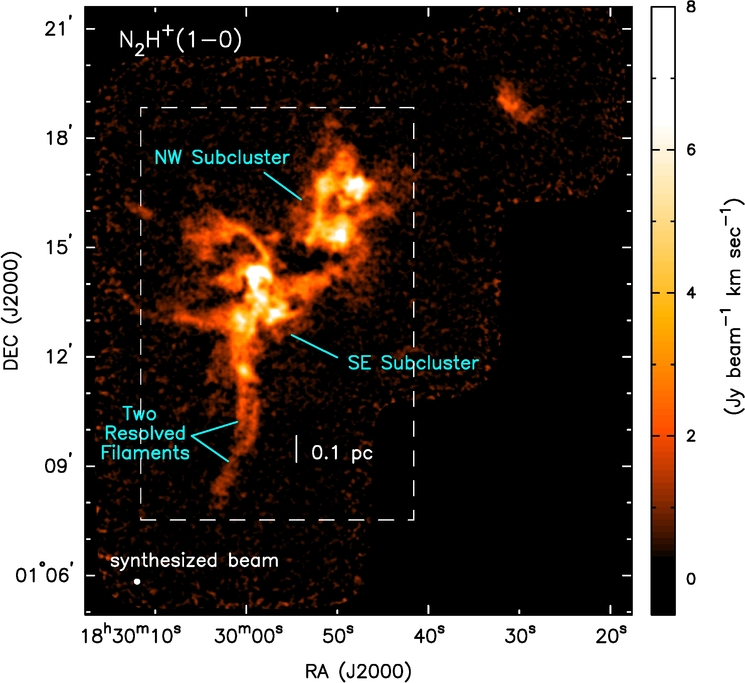}
\caption{
\footnotesize
The integrated intensity map of \nh \ in Serpens Main. 
The map is integrated over the strongest hyperfine line from 5.02 to 10.83 km~s$^{-1}$.
The noise level is 0.24 Jy~beam$^{-1}$~km~s$^{-1}$.
The beam size is $7.7\arcsec \times 7.0\arcsec$ as shown in the bottom left corner.
The white, dash rectangle indicates the region we plot in most of the subsequent figures.   
}
\label{fig:serpens_n2hp}
\end{center}
\end{figure}

\begin{figure}[h]
\begin{center}
\begin{tabular}{c@{\hspace{0.15 in}}c}
\includegraphics[scale=1.56,angle=0]{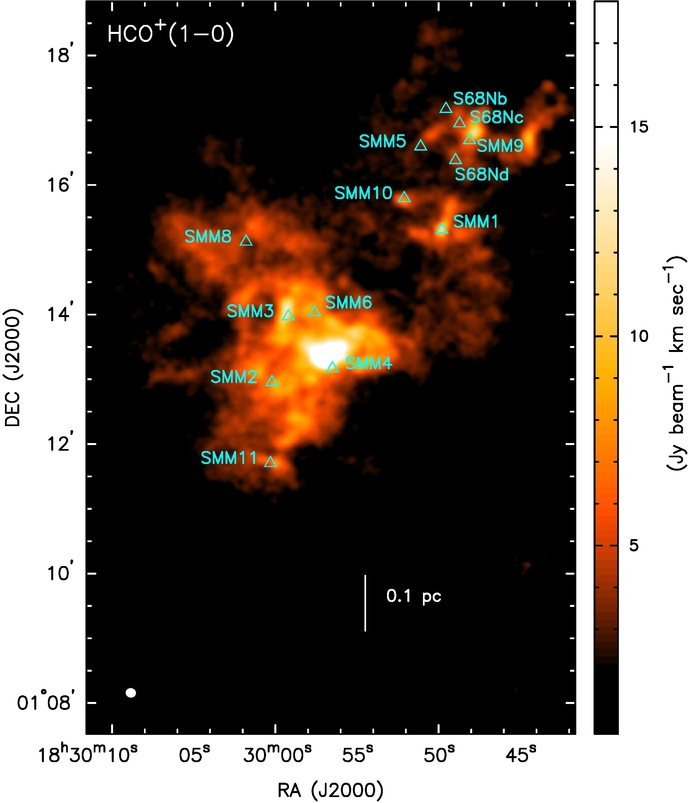} &
\includegraphics[scale=1.56,angle=0]{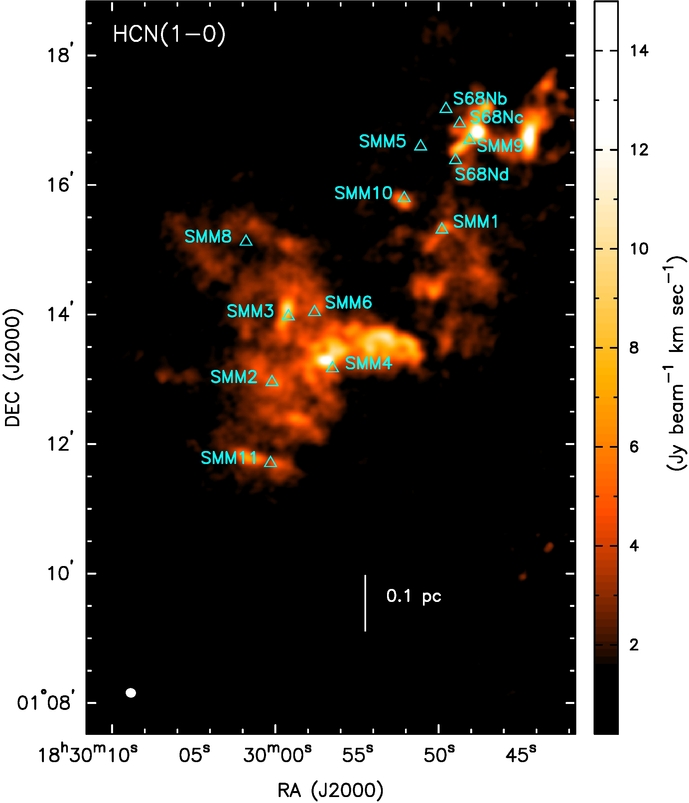}
\end{tabular}
\end{center}
\caption{\footnotesize
The integrated intensity maps of \hcop \ (left panel; $\sigma=0.27$ Jy~beam$^{-1}$~km~s$^{-1}$) and \hcn \ (right panel; $\sigma=0.22$ Jy~beam$^{-1}$~km~s$^{-1}$) in Serpens Main.
The maps are integrated over a velocity range of 4.07 km~s$^{-1}$ to 11.94 km s$^{-1}$ for \hcop, and 4.04 km~s$^{-1}$ to 10.98 km~s$^{-1}$ for \hcn.  The cyan triangles are submillimeter sources from \citet{1999MNRAS.309..141D}.
The synthesized beam is drawn as the white circle at the bottom left corner.
}
\label{fig:serpens_hcop}
\end{figure}

\subsection{Global Gas Morphology from N$_{2}$H$^{+}$, HCO$^{+}$, and HCN}
\label{sect:mom0}

Figure~\ref{fig:serpens_n2hp} shows the integrated intensity map from the strongest hyperfine line in the \nh \ observations. 
There are two main features shown in the map.
First, the peaks in the integrated intensity map are concentrated on the NW and SE subclusters. 
The \nh \ gas peaks are distributed over the extent of the NW subcluster, 
while in the SE subcluster the emission peaks are more concentrated in the central region.  
Second, prominent filamentary structures are observed in the regions that were not well resolved by previous observations, especially in the SE subcluster.
In particular, the southern filament, which was not well resolved previously \citep[e.g.,][]{1999MNRAS.309..141D,2010A&A...518L.102A}, has been resolved into two narrower filaments for the first time.

The region in the northwest corner of Figure~\ref{fig:serpens_n2hp} (centering at $\alpha_{J2000}$=18$^{h}$29$^{m}$31.8$^{s}$, $\delta_{J2000}$=01\arcdeg18\arcmin44.5\arcsec) shows \nh \ gas emission which peaks at $\sim 2$ Jy~beam$^{-1}$ in the strongest hyperfine line, about 50\% of the typical \nh \ gas peaks ($\sim4$ Jy~beam$^{-1}$) in the central regions of the NW and SE subclusters.  
The \hcop \ emission in this region only shows in 3 channels with a weak peak of $\sim1$ Jy~beam$^{-1}$,
20\% of a typical peak of $\sim 5$ Jy~beam$^{-1}$ in the central regions of the two subclusters.
The peak intensity in the strongest \hcn \ hyperfine line at this region is about 1.2 Jy~beam$^{-1}$, 40\% of the typical \hcn \ emission ($\sim 3$ Jy~beam$^{-1}$) in the two subclusters.  
In general, this region shows weak emission compared to the central regions in the two subclusters for all the three molecules.
This paper will mainly focus on the two subclusters in the following sections.  

Figure \ref{fig:serpens_hcop} shows the \hcop \ and \hcn \ integrated intensity maps
 of the two subclusters.
The two molecular lines exhibit similar emission distributions, which are more extended than the \nh \ emission; however, 
the filamentary structures are not prominent. In particular, the long southern filament is not obvious in the integrated intensity maps but weak narrow lines are present along some parts of the filament. 
The strongest \hcop \ and \hcn \ emission peaks are associated with a few of the eleven submillimeter cores (SMM1 - SMM11) identified by \citet{1999MNRAS.309..141D}.
The \nh \ emission peaks are also associated with several submillimeter cores (see discussions in Sect.~\ref{sect:leaf}).

The observed differences in the emission distributions of \nh \ , \hcop \ , and \hcn \ are likely dominated by differences in their abundance distributions due to chemical effects. \nh \ favors cold dense gas where CO becomes depleted because chemical reaction with CO is a major destroyer of \nh \ \citep{2001ApJ...557..209B,2002ApJ...570L.101B}.
\hcop \ favors environments with significant ionization fraction, which tends to be warm gas \citep{2010A&A...520A..20G}. 
\hcn \ is prominent in warm gas and depletes with CO in cold dense gas \citep{2006A&A...455..577T}. 
All three molecular transitions have critical densities around few times 10$^5$~cm$^{-3}$ \citep{1999ARA&A..37..311E}, so their emissivity peaks in dense gas. 
However, all three are resonance transitions (the lower level is the ground state), so lower density gas can emit and absorb in these $J=1 \rightarrow 0$ lines. For \hcop \ and \hcn, the fact that their abundance remains high or increases in the lower density gas means that their emission can trace gas from density 10$^4$ to $> 10^{6}$~cm$^{-3}$, and the lower density gas can absorb away emission from high density gas. 
The chemical selectivity of \nh \ favors dense gas which minimizes the contributions from, and impact of, lower density gas.
The focus of the analysis in this paper will be on the \nh \ emission because, 
of the three lines observed, it is the best dense gas tracer.

\begin{figure}
\begin{center}
\includegraphics[scale=2.1,angle=270]{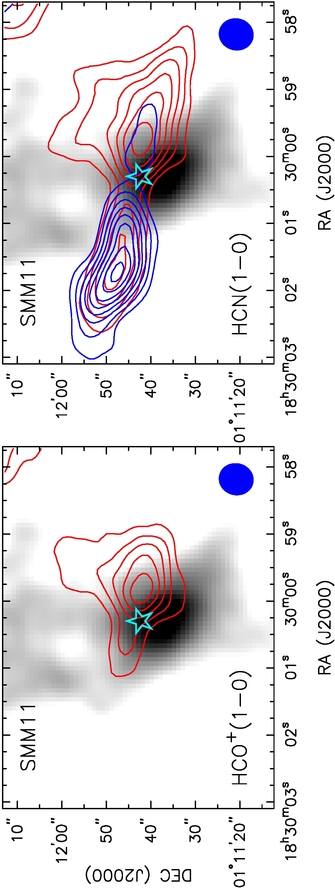}
\end{center}
\caption{\footnotesize
Integrated intensity map of the SMM11 outflow from \hcop \ (\textit{left panel}) and \hcn \ (\textit{right panel}).
Blue (red) contours represent the blueshifted (redshifted) lobe.
The grey scale is the integrated intensity map of \nh.
The star symbol marks the position of SMM11.
The synthesized beam is drawn as the blue circle at the bottom right corner.
\textit{Left panel:} The red lobe is integrated from 8.5 km~s$^{-1}$ to 12.11 km~s$^{-1}$.  Contours: 5, 7, 9, 11, 13$\times \sigma$ ($\sigma=0.21$ Jy~beam$^{-1}$~km~s$^{-1}$).  \hcop \ did not show a blue lobe.
\textit{Right panel:} The red lobe is integrated from 9.33 km~s$^{-1}$ to 10.81 km~s$^{-1}$.  Contours: 5, 7, 9, 11, 13, 15$\times \sigma$ ($\sigma=0.13$ Jy~beam$^{-1}$~km~s$^{-1}$).
The blue lobe is integrated from 2.72 km~s$^{-1}$ to 6.85 km~s$^{-1}$.  Contours: 6, 9, 12, 15, 18, 21, 24, 27$\times \sigma$ ($\sigma = 0.18$ Jy~beam$^{-1}$~km~s$^{-1}$).
}
\label{fig:outflow}
\end{figure}

\hcop \ and \hcn \ have been considerably used in detecting outflow activity \citep[e.g.,][]{2013ApJ...778...72F}. 
Several sources including SMM1, SMM2, SMM3, SMM4, SMM8, and SMM9 have been suggested to be associated with outflows from observations of CO and its isotopes \citep{1995A&A...298..594W,2010MNRAS.409.1412G}.
We show an example of outflow associated with SMM11 in Figure~\ref{fig:outflow}.
The left panel shows the integrated intensity map of the \hcop \ redshifted emission from SMM11 (the star symbol), and the right panel shows the integrated intensity map of the \hcn \ redshifted and blueshifted emission.
For \hcn, the outflow is identified based on the main component of the three hyperfine lines.  
\hcn \ clearly traces a collimated outflow powered by SMM11 in the center. 
The outflow is in a northeast-southwest direction with a P.A.\ of about 75\arcdeg. 
This direction is nearly perpendicular to the southern filament.  
The red lobe has extended emission to the north; a similar morphology in the red lobe is also seen in the \hcop \ map.
However, the blue lobe is not detected with \hcop.  
An \nh \ peak shown in the grey-scale image is very close to SMM11.  
The relation between the \nh \ emission peak and SMM11 is not clear since the emission enhancement in \nh \ might be due to the two overlapping filaments (Sect.~\ref{sect:kinematics}).  
Also, the small offset between the \nh \ emission peak and the source may be due to chemical effects from the central heating of the source \citep[e.g.,][]{2011A&A...525A.141B,2013A&A...560A..41L}.

\subsection{Global Gas Kinematics from N$_{2}$H$^{+}$}
\label{sect:kinematics}

\begin{figure}
\begin{center}
\includegraphics[scale=1.44]{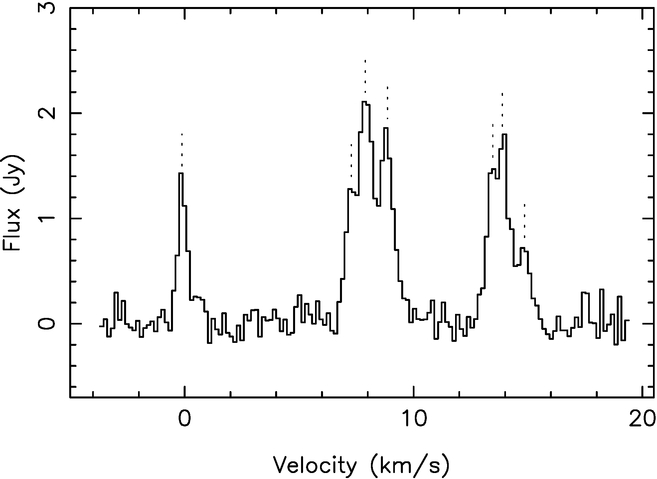}
\includegraphics[scale=1.44]{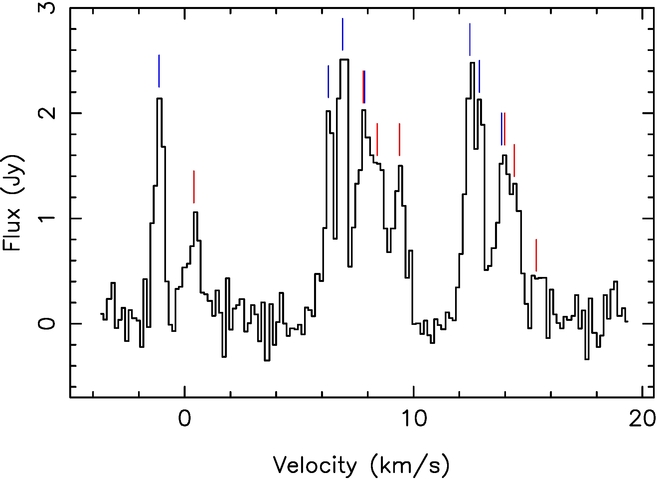}
\caption{\footnotesize
\textit{Left Panel:} \nh \ line profile averaged over the area of a synthesized beam (7.7\arcsec \ by 7.0\arcsec) centered on the position (RA = 18:29:57.8, Dec = +01:14:42.3), which shows a typical cloud spectrum with seven hyperfine components.  The seven hyperfine components are marked with the dashed lines ($\text{V}_{\text{lsr}}=8.14$ km~s$^{-1}$).
\textit{Right Panel:}
\nh \ line profile showing an example of a two-velocity component spectrum at the location with the star symbol in Fig.\ \ref{fig:serpens_v}.
The two components are clearest in the isolated hyperfine component near 0~km~s$^{-1}$.
The hyperfine lines with the two velocity components ($\text{V}_{\text{lsr}}=7.14$ and $8.66$ km~s$^{-1}$) are indicated as the blue and red lines, respectively.
}
\label{fig:multiv}
\end{center}
\end{figure}

\begin{figure}[h]
\begin{center}
\begin{tabular}{c@{\hspace{0.1 in}}c}
\includegraphics[scale=1.56,angle=0]{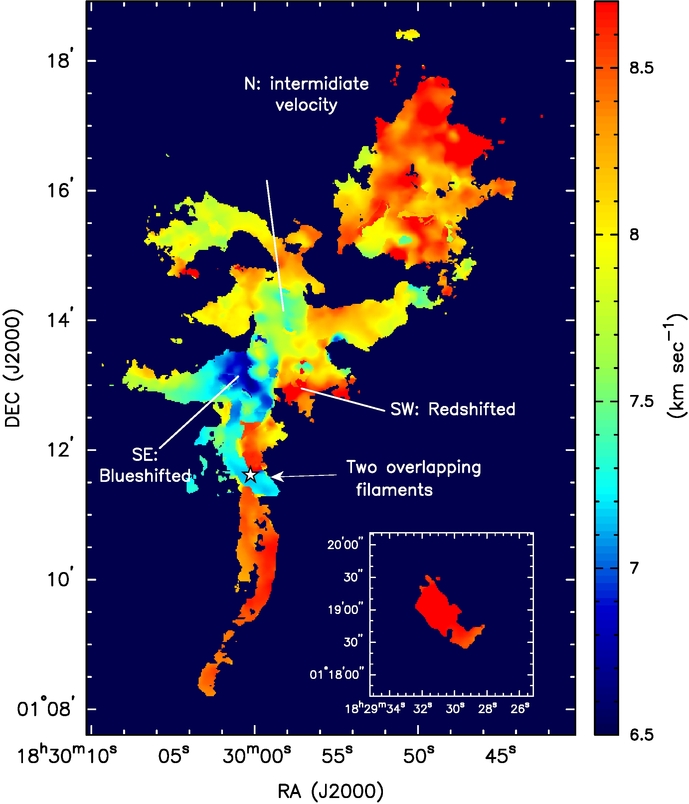} &
\includegraphics[scale=1.56,angle=0]{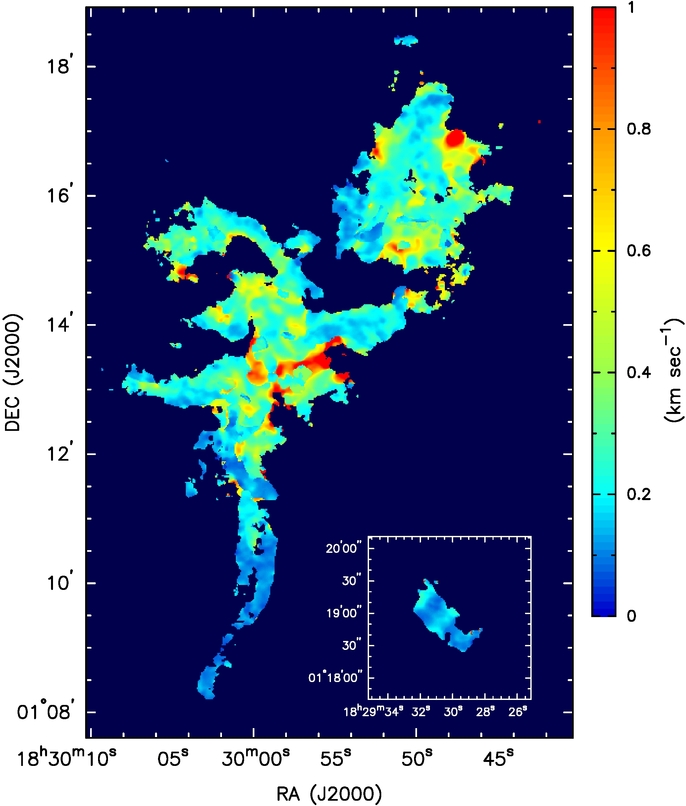}
\end{tabular}
\end{center}
\caption{\footnotesize
\textit{Left panel:} The centroid velocity map from \nh \ spectral line fitting.  The NW subcluster shows a relatively uniform structure in velocity while the SE subcluster shows more complicated velocity fields including a blueshifted region in the southeast, a redshifted region in the southwest, and a region with intermediate velocity in the north.
The star symbol indicates the location of SMM11.
Two filaments with different velocities overlap near the position of SMM11.
\textit{Right panel:} The velocity dispersion map from \nh \ spectral line fitting.
Velocity dispersions show subsonic to sonic gas motions in 60\% of the region.
The central region in the SE subcluster shows larger velocity dispersions than the surrounding filaments. 
The subplots in the lower right in both panels show the northwest corner in Fig.~\ref{fig:serpens_n2hp}. 
}
\label{fig:serpens_v}
\end{figure}

Since the \nh \ gas emission is less contaminated by lower density gas and by outflows like \hcop \ and \hcn \ (Sect.~\ref{sect:mom0}), we use \nh \ as the main probe for studying dense gas kinematics.
We obtained the centroid velocities ($\text{V}_{\text{lsr}}$) and velocity dispersions ($\sigma$) of \nh \ by fitting all the seven hyperfine lines simultaneously using Gaussian profiles on a pixel-by-pixel basis.
The fitting method and the detailed algorithm are described in Paper I.
We have incorporated the pixels with (1) a peak signal-to-noise ratio larger than 5~in the spectrum, and 
(2) a signal-to-noise ratio larger than 4~in the integrated intensity map.  
Pixels below the threshold are blanked in the centroid velocity and velocity dispersion maps.
The left panel in Figure~\ref{fig:multiv} shows a typical \nh \ line profile with the seven hyperfine components and fitted centroid velocities.

Multiple velocity components are observed along several lines of sight in both subclusters; it is most evident in the isolated component in the hyperfine structure.
We performed two-component line fitting in the locations with two velocity components along the line of sight with the inclusion of a second set of hyperfine lines.  
The stronger component in the peak intensity with the two-component fitting was chosen as the primary component shown in the kinematic maps.  
The most obvious example of two velocity components is at the location of SMM11. 
The right panel in Figure \ref{fig:multiv} shows the spectrum at that location; 
the $\text{V}_{\text{lsr}}$ of the two components are 7.14 km~$s^{-1}$ and 8.66 km~s$^{-1}$. 
We do not see a correlation between the locations of two-velocity components and those of YSOs or filaments.

The left panel in Figure~\ref{fig:serpens_v} shows the centroid velocity map from the fitting of the \nh \ data.
As seen in the centroid velocity map,
the velocity fields in the two subclusters are distinct.
The NW subcluster presents relatively uniform velocity fields,
while the SE subcluster has a more complicated velocity pattern.
Most of the gas in the NW subcluster has centroid velocities between 8.0 and 8.5 km~s$^{-1}$.
Most of the SE subcluster shows more blueshifted centroid velocities compared to the NW subcluster. 
The gas kinematics near SMM11 (shown as the star symbol in the centroid velocity map) reveal two filamentary structures with different velocities (V$_{\text{lsr}} \sim$7.14 and 8.66 km~$^{-1}$, respectively), suggesting that these two filaments are distinct and overlapped along the line of sight at the position of SMM11.
The northwest corner in Fig.~\ref{fig:serpens_n2hp} has centroid velocities around 8.5 km~s$^{-1}$ as shown in the subplot.

The center of the SE subcluster has three main regions with different velocities: th blueshifted area in the southeast, the redshifted area in the southwest, and the area in the north with intermediate velocities.  
The two crossing filaments in the south appear to have blueshifted and redshifted velocities similar to the southeast blueshifted and southwest redshifted areas, respectively, 
suggesting that the southeast area may be connected to the blueshifted filament, and the southwest area possibly connects to the redshifted filament. 

The right panel in Figure~\ref{fig:serpens_v} shows the velocity dispersion map, also obtained from the \nh \ fitting.
The velocity dispersions have a median value of 0.24 km~s$^{-1}$ and a mean value of 0.30 km~s$^{-1}$.  
Assuming a gas temperature of 20 K for the overall cloud \citep{2000ApJ...536..845M}, which suggests
an isothermal sound speed of $\sim$ 0.27 km s$^{-1}$, 
about 60\% of the region display a subsonic to sonic level velocity distribution along the line of sight.
In the SE subcluster, the central region appears to have larger velocity dispersions (typically $>$0.5 km~s$^{-1}$) than the surrounding filaments (typically $<0.2$ km~s$^{-1}$).  
The southern filaments are especially quiescent with velocity dispersions $\sim$ 0.1 km s$^{-1}$.
The northwest corner in Fig.~\ref{fig:serpens_n2hp} has subsonic velocity dispersions as shown in the subplot.

\begin{figure}[h!]
\begin{center}
\includegraphics[scale=1.0]{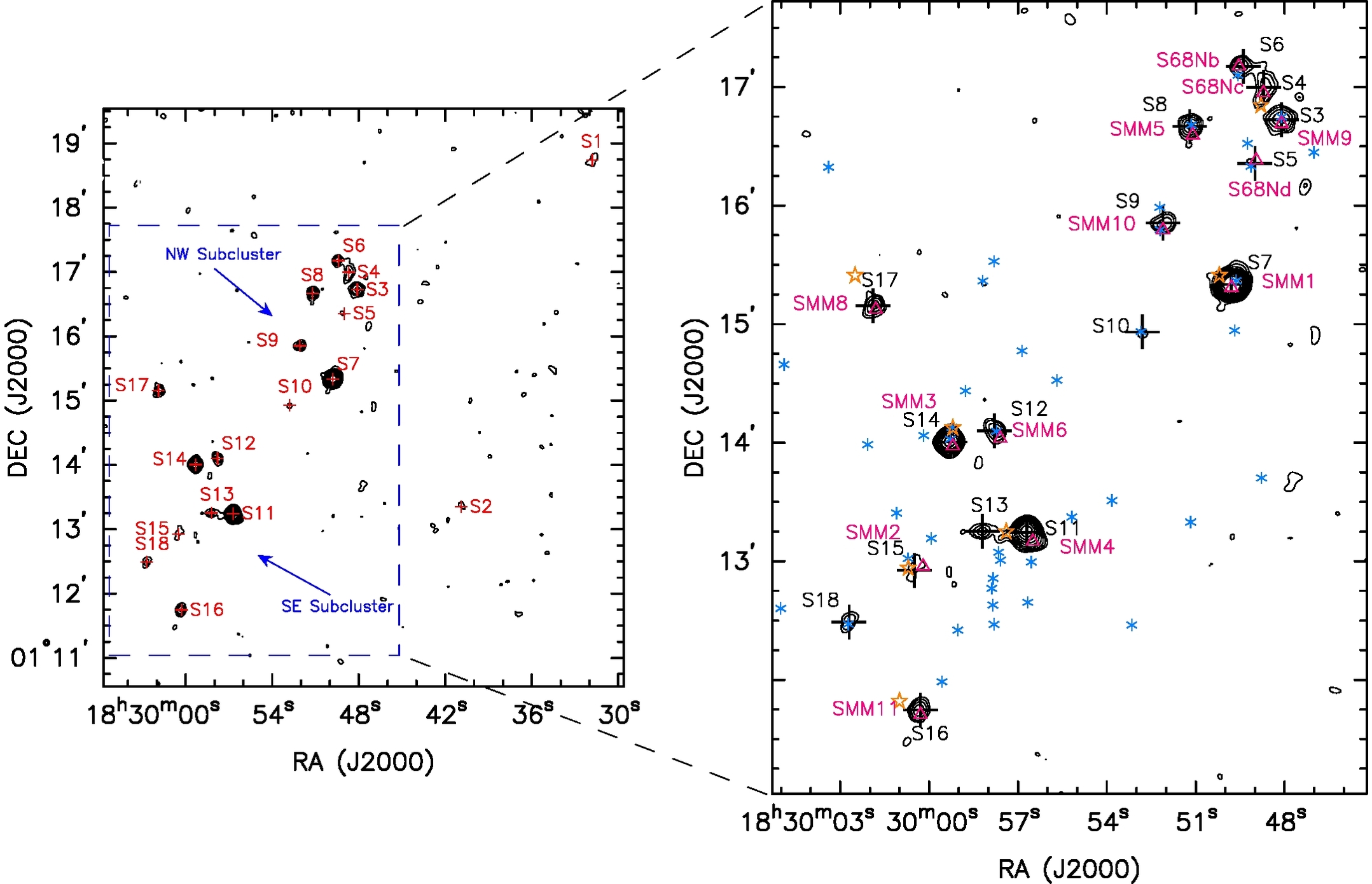}
\caption{\footnotesize
\textit{Left panel}: The 3 mm continuum map of the Serpens Main region.  The contour levels are 3, 5, 7, 9, 11, 15, 20, 30, 40, 50, 60 $\times \sigma$ ($1\sigma=1.5$ mJy beam$^{-1}$).  Eighteen sources (S1 - S17) are identified in the map and most of them are concentrated in the NW and SE subclusters.
\textit{Right panel}:
Comparison of continuum sources with previous results.  The contours show the continuum map at 3~mm from this work.  Contour levels are 3, 4.2, 6, 8.4, 12, 16.8, 24, 33.6, and 48 times the sigma level ($\sigma=1.5$ mJy beam$^{-1}$).
The black crosses show the fitted positions of the 3~mm sources from this work.
The magenta triangles are the submillimeter sources identified with the SCUBA 450 $\um$ and 850 $\um$ observations from \citet{1999MNRAS.309..141D}.
The orange stars are the 1~mm sources identified with Bolocam observations from \citet{2007ApJ...666..982E}.
The blue asterisks are the YSOs identified by \citet{2009ApJS..181..321E} with the \textit{Spitzer} IRAC and 2MASS observations.
}
\label{fig:conticores}
\end{center}
\end{figure}

\subsection{Continuum Sources}
\label{sect:conti}

\begin{deluxetable}{lcc c c c r@{ $\pm$ }l  c r@{ $\pm$ }lc}
\tabletypesize{\tiny}
\tablewidth{0pc}
\tablecaption{Core Properties from CARMA 3~mm Continuum Data
\label{tbl:conti}}
\tablehead{
\colhead{Source} & \colhead{RA\tablenotemark{a}} &
\colhead{DEC\tablenotemark{a}} & \colhead{Major $\times$ Minor\tablenotemark{a}} & \colhead{Size\tablenotemark{b}} & \colhead{Aspect\tablenotemark{c}} & \multicolumn{2}{c}{Peak Brightness\tablenotemark{d}} &\colhead{Total Flux Density} & \multicolumn{2}{c}{Mass} & \colhead{Other} \\
\colhead{} &\colhead{(J2000)} &
\colhead{(J2000)} & \colhead{Axes (\arcsec)} & \colhead{(AU)} & \colhead{Ratio} & \multicolumn{2}{c}{(mJy beam$^{-1}$)} & \colhead{(mJy)} & \multicolumn{2}{c}{(M$_{\sun}$)} & \colhead{Identifier}
}
\startdata
S1 & 18:29:31.8 & +01:18:44.5 & 14.7 $\times$ 6.2 & 4000 & 2.4  & 7.6&1.1 & 20.5 $\pm$ 1.4 & 1.15&0.08 & \\
S2 & 18:29:40.9 & +01:13:21.1 & ... & ... & ...  & 6.1&1.5 & ...  & 0.34&0.08\tablenotemark{e} & \\
S3 & 18:29:48.1 & +01:16:43.5 & 8.9 $\times$ 7.6 & 3400 & 1.2  & 18.7&1.7 & 42.6 $\pm$ 1.9 &  2.38&0.11 & SMM9 \\
S4 & 18:29:48.7 & +01:16:59.9 & 26.7 $\times$ 8.6 & 6300 & 3.1 & 9.2&2.0 & 54.4 $\pm$ 4.2 &  3.04&0.23 & S68Nc \\
S5 & 18:29:49.0 & +01:16:21.2 & ... & ... & ... & 5.8&1.5 & ... & 0.32&0.08 & S68Nd \\
S6 & 18:29:49.4 & +01:17:10.4 & 6.7 $\times$ 5.1 & 2400 & 1.3  & 13.4&1.7 & 22.1 $\pm$ 1.4 &  1.24&0.08 & S68Nb \\
S7 & 18:29:49.8 & +01:15:20.1 & 4.5 $\times$ 4.1 & 1800 & 1.1  & 149.3&3.9 & 201.6 $\pm$ 2.3 &  11.27&0.13 & SMM1 \\
S8 & 18:29:51.2 & +01:16:40.0 & 7.1 $\times$ 4.6 & 2400 & 1.5  & 18.9&2.2 & 31.2 $\pm$ 1.7  & 1.74&0.10 & SMM5 \\
S9 & 18:29:52.1 & +01:15:51.2 & 7.2 $\times$ 3.5 & 2100 & 2.1  & 12.8&1.2 & 19.8 $\pm$ 0.8 &  1.11&0.04 & SMM10 \\
S10 & 18:29:52.8 & +01:14:56.0 & ... & ... & ...  & 6.4&1.5 & ...  &  0.36&0.08\tablenotemark{e} &  \\
S11 & 18:29:56.7 & +01:13:14.6 & 6.5 $\times$ 4.5 & 2200 & 1.4  & 67.0&4.1 & 106.0 $\pm$ 3.0 &  5.92&0.17 & SMM4 \\
S12 & 18:29:57.8 & +01:14:06.0 & 7.7 $\times$ 3.8 & 2200 & 2.0  & 13.4&1.5 & 22.2 $\pm$ 1.1 & 1.24&0.06 & SMM6 \\
S13 & 18:29:58.2 & +01:13:15.3 & 9.6 $\times$ 2.3 & 2000 & 4.2  & 10.2&1.1 & 17.6 $\pm$ 0.7 & 0.98&0.04 \\
S14 & 18:29:59.3 & +01:14:00.4 & 4.3 $\times$ 3.8 & 1700 & 1.1  & 47.7&1.8 & 62.5 $\pm$ 1.0 & 3.49&0.06 & SMM3 \\
S15 & 18:30:00.5 & +01:12:55.4 & ... & ... & ...  & 6.3&1.5 & ... & 0.35&0.08\tablenotemark{e} & SMM2 \\
S16 & 18:30:00.3 & +01:11:44.7 & 6.6 $\times$ 2.9 & 1800 & 2.3  & 16.5&1.2 & 24.1 $\pm$ 0.7  & 1.35&0.04 & SMM11 \\
S17 & 18:30:01.9 & +01:15:09.2 & 7.4 $\times$ 3.8 & 2200 & 1.9  & 14.3&1.7 & 23.2 $\pm$ 1.2  & 1.30&0.07 & SMM8 \\
S18 & 18:30:02.7 & +01:12:29.3 & point source & ... & ... &  9.1&1.0 & 14.5 $\pm$ 1.3 &  0.81&0.07 \\
\enddata
\tablenotetext{a}{The positions, major and minor axes are estimated with the MIRIAD IMFIT task.
The major and minor axes here are the deconvolved sizes of the two axes. Errors on deconvolved sizes are $\pm$ 1$^{\prime\prime}$.  We do not report the major and minor axes for sources $<$ 5$\sigma$.}
\tablenotetext{b}{Geometric sizes defined as $\sqrt{\text{Major axis} \times \text{Minor axis}}$}.
\tablenotetext{c}{Aspect ratio is calculated as the major axis divided by the minor axis.}
\tablenotetext{d}{For sources $> 5\sigma$, peak brightness and total flux densities are estimated using the MIRIAD IMFIT task.
For sources $< 5\sigma$, peak brightness are estimated using the MIRIAD HISTO task (we do not report total flux densities for sources $< 5\sigma$).}
\tablenotetext{e}{Masses denote lower limits for sources $<$ 5$\sigma$.}
\end{deluxetable}

We identified 18 continuum sources at 3~mm (S1 - S18: Table~\ref{tbl:conti}) as shown in Figure~\ref{fig:conticores} (left panel).
The sources were identified based on two criteria:
(1) peak brightness $\ge$ $5\sigma$ or
(2) peak brightness $\ge$ $3\sigma$ and an association with an emission peak in at least one of the following bands:
\textit{Spitzer} 8, 24, 70 $\um$, \textit{Herschel} 160, 250, 350, 500 $\um$, where the
peak must be within half of the largest beam among the bands showing detections.
The sources are mostly concentrated in the NW and SE subclusters, consistent with previous results \citep[e.g.,][]{1999MNRAS.309..141D,2000ApJ...537..891W}.
S1 and S2 are in more isolated positions.
There are 8 sources in the NW subcluster, and 8 sources in the SE subcluster.
\citet{1998ApJ...508L..91T} reported 32 continuum sources above 4.0 mJy~beam$^{\text{-1}}$ with a sensitivity of 0.9 mJy~beam$^{\text{-1}}$.  
All of our sources are detected by \citet{1998ApJ...508L..91T} except S1 and S2 (outside their mapping area).
However, we could not confirm most of their sources with peaks of $\sim4.0$ mJy~beam$^{\text{-1}}$, which corresponds to 2.5 to 3 $\sigma$ levels in our map.

The positions, major and minor axes, aspect ratios, peak brightness, and total flux densities 
of these sources are given in Table~\ref{tbl:conti}. 
For sources with $>$ 5$\sigma$ detections, the quantities were determined by Gaussian fits to the emission. 
For $<$ 5$\sigma$ sources, we report the peak brightness and omit major, minor axes and total flux densities.
S7 (in the NW subcluster) has the highest brightness among all the sources
and S11 (in the SE subcluster) has the second highest brightness.  
The geometric sizes range from $\sim 1700$ AU to $\sim 6300$ AU with a mean value of 2700 AU.
S4 is particularly extended and may be multiple sources.
The aspect ratios range from 1.1 to 4.2 with an averaged value of 2.0,
suggesting that the sources are elongated and not spherical.

The right panel in Figure~\ref{fig:conticores} shows a comparison of continuum sources from previous observations at multiple wavelengths including the SCUBA 850 $\um$ submillimeter sources from \citet{1999MNRAS.309..141D},
the Bolocam 1~mm sources from \citet{2007ApJ...666..982E}, and the \textit{Spitzer} YSOs from \citet{2009ApJS..181..321E}.
\citet{1999MNRAS.309..141D} identified 11 submillimeter sources (SMM1 - SMM11) and three ``sub-clumps" (a, b, and c) associated with SMM9 (S68N), which were resolved by high angular resolution observations carried out by \citet{2000ApJ...537..891W} using BIMA at 3~mm (S68Na - S68Nd).
All of the submillimeter sources have correspondences to 3~mm sources, except for S68Nd.
The majority of these submillimeter sources are associated with Class 0/I objects \citep{2007ApJ...669..493W,2010ApJ...710.1247S} with outflow activities \citep[e.g.,][]{1999MNRAS.309..141D,2010MNRAS.409.1412G}.
Despite the good correlation between the submillimeter and 3~mm sources, not all the submillimeter sources are associated with the 1~mm sources; this is possibly due to the limitations from the sensitivity and the angular resolution (30\arcsec) in the Bolocam observations.
Only three sources (S7, S14, and S15) have all the counterparts in \textit{Spitzer} YSOs, the submillimeter cores, the 1~mm cores, and the 3~mm cores.

We calculate the masses of the continuum sources with the relation:
\begin{equation}
M=\frac{d^{2}F_{\nu}}{B(T_{d})\kappa_{\nu}},
\end{equation}
where  $F_{\nu}$, $d$, $\kappa_{\nu}$, and $B_{\nu}(T_d)$, are respectively the total observed flux density, distance, grain opacity,
and blackbody intensity at dust temperature, $T_d$.
We adopt $T_d$ = 20 K; the filament temperatures (Sect.~\ref{sect:filament}) are 12-14 K, so we assume the average
dust temperature near the protostar is slightly higher.
We estimate $\kappa_{\nu}$ at observed wavelength using $\kappa_{\nu}$ = 0.1($\nu$/10$^{3}$ GHz)$^{\beta}$ cm$^{2}$~g$^{-1}$ \citep{1990AJ.....99..924B} with an assumed dust-to-gas ratio of 100. 
Assuming $\beta \sim 1.5$,
we obtain $\kappa_{\nu}$ = 0.0027 $\text{cm}^2~\text{g}^{-1}$ at 3.3~mm.
The masses of the sources $<$ 5$\sigma$ are calculated from the peak brightness and represent lower limits.
The resulting masses are summarized in Table~\ref{tbl:conti}.
Most of the sources have masses ranging from 0.3 to 3.5 M$_{\sun}$. 
The two brightest sources, S7 and S11, have masses of 11.3 M$_{\sun}$ and 5.9 M$_{\sun}$, respectively.  Uncertainties in $\kappa_{\nu}$, dust temperature, $\beta$, and distance imply that the masses have systematic uncertainties at the factor of two level.
\clearpage

\section{Dendrogram Analysis: Characterizing Hierarchical Structures from Small to Large Scales}
\label{sect:dendro}

We performed a dendrogram analysis on the \nh \ data to study the
structure of dense gas in Serpens Main following the methodology presented in
\citep{2008ApJ...679.1338R}.  Dendrogram analysis has been used to
investigate hierarchical, cloud-to-core gas structures in several
recent works
\citep[e.g.,][]{2009Natur.457...63G,2010ApJ...712.1137K,2013ApJ...770..141B}.
We summarize the basic concepts here.  A dendrogram is a tree diagram
that characterizes how and where structures surrounding local maxima
in position-position-velocity (PPV) space merge. Structures grow in
volume with decreasing flux density level until they encounter adjacent
structures. Local maxima in intensity become the tips of leaves of the
tree if they pass criteria used to suppress noise features; the
most important criterion is that a local maximum must exceed an
intensity threshold above its merge level with
another local maximum. The merge level, defined as the isocontour
which encircles two or more leaves, creates a branch which grows in
volume until it encounters another leaf or branch. Joining with a leaf or
branch creates a lower level branch which can then repeat the cycle of
growth and merger until the chosen stopping flux level is reached. 
We utilize the new ``non-binary" merging algorithm described by Paper I.
In the ``binary" merging algorithm leafs and branches can only combine in pairs, which creates artificial branching structures; 
the ``non-binary" algorithm allows merging of multiple leafs and branches as dictated by the chosen minimum rms step between mergers.

\begin{figure}[h!]
\begin{center}
\includegraphics[scale=0.35]{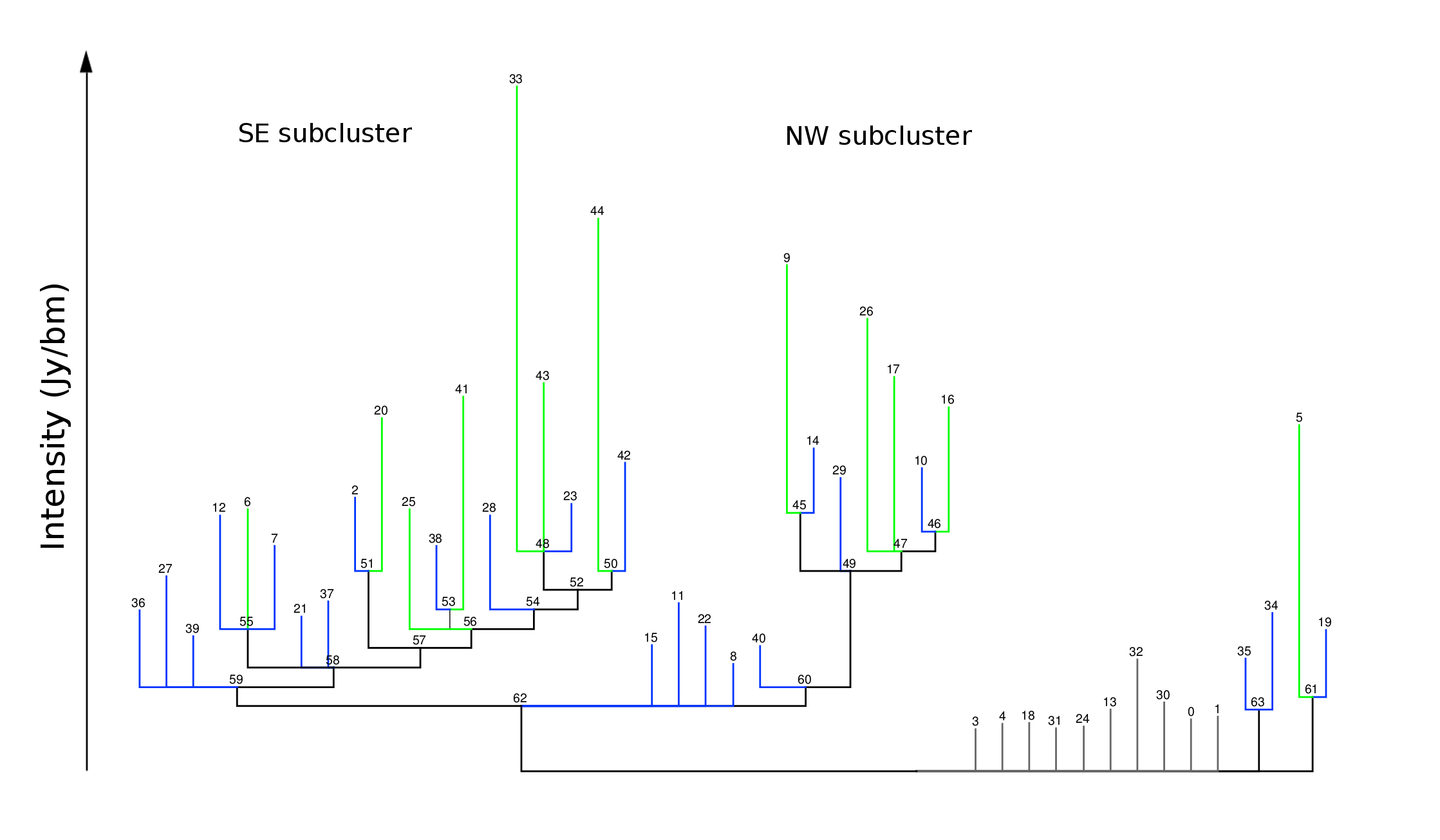}
\caption{\footnotesize The non-binary dendrogram for Serpens Main.  The vertical axis indicates intensity in Jy~beam$^{-1}$.
High-contrast leaves peak at least 6$\sigma$ in intensity above their nearest branch, and are colored green, while low-contrast leaves peak below 6$\sigma$, and are colored blue.
Leaves that grow directly from the base of the tree without any branching (``sprouts") are colored grey.
There are 12 high-contrast leaves, 33 low-contrast leaves (including sprouts), and 19 branches.}
\label{fig:tree}
\end{center}
\end{figure}

To optimize the structure identification with a dendrogram, we binned
the \nh \ data by 2 channels ($\sigma=0.11$ Jy beam$^{-1}$) for
better signal-to-noise ratios with the kinematic information
preserved.  The major parameters in the algorithm are: 1) 
a 3-dimensional spatial-velocity size for identifying separate local maxima, 
with $10\arcsec \times 10\arcsec$ spatial dimensions and a 3-channel velocity dimension 
(with single channel width of 0.314 km~s$^{-1}$ for the  binned data cube), 
2) the 2$\sigma$ intensity threshold parameter for culling local maxima, and 3)
a minimum of 3 synthesized beams of spatial-velocity pixels belonging to a leaf for it to be considered real.
An initial masking was performed based on a $4\sigma$ sensitivity level with an
expansion to adjacent pixels of $2.5\sigma$.  We used the isolated
component in the \nh \ hyperfine line structure to avoid mixture of
emission from blending lines in the dendrogram analysis.

\subsection{The Dendrogram Technique}
\label{sect:leaf}

\begin{figure}[t!]
\begin{center}
\includegraphics[scale=2.1]{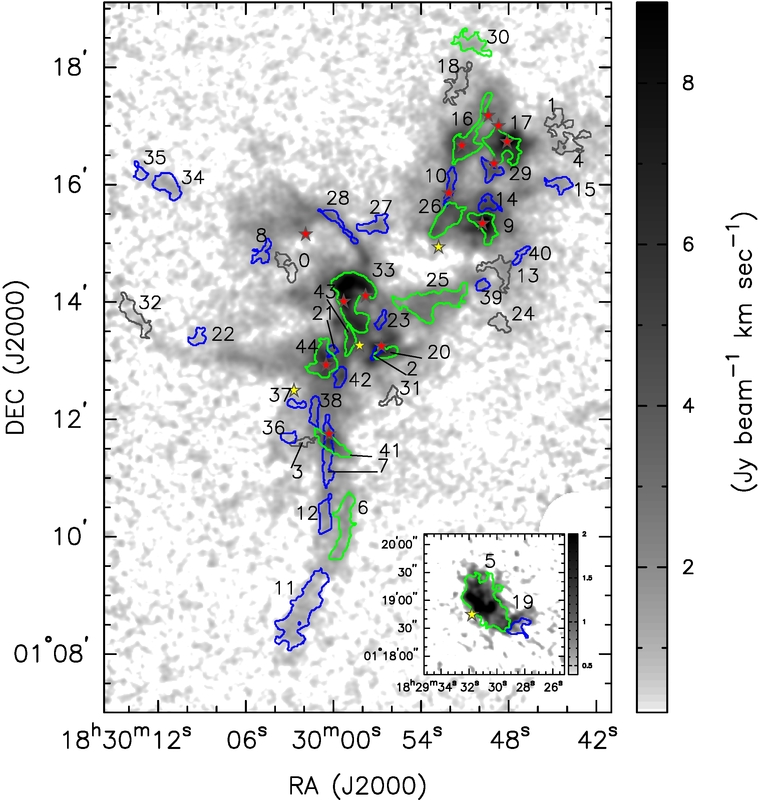}
\caption{\footnotesize Leaves in the dendrogram overlaid on the \nh \ integrated intensity map (the greyscale image).  Green: high-contrast leaves (the contrast larger than 6$\sigma$).  Blue: low-contrast leaves (the contrast smaller than 6$\sigma$).  Grey: sprouts.  Yellow stars indicate the 3~mm continuum sources from the CLASSy observations. Red stars indicate the SMM cores \citep{1999MNRAS.309..141D}.  All the SMM cores have counterparts at 3~mm (see Sect.~\ref{sect:conti}).
The subplot in the lower right shows the northwest corner in Fig.~\ref{fig:serpens_n2hp}.
}
\label{fig:leaf}
\end{center}
\end{figure}

Figure~\ref{fig:tree} shows the resulting tree from the non-binary dendrogram analysis.
The tree structure beginning at branch 59 and extending upward represents the SE subcluster, while the structure beginning at branch 60 represents the NW subcluster.
To better compare the intensities of the leaves and their properties,
we differentiate stronger leaves from weaker leaves using a contrast criterion: ``high-contrast" leaves have peak intensities at least 6$\sigma$ in intensity above the the level of their nearest branch, while ``low-contrast" leaves have peak intensities below $6\sigma$ (contrast values are listed in Table~\ref{tbl:tree}).
There are 12 high-contrast leaves and 33 low-contrast leaves.
We further define ``sprouts" as the leaves directly growing out from the base tree (colored as grey in Fig.~\ref{fig:tree}).
All the sprouts are low-contrast leaves in Serpens Main.  
Figure~\ref{fig:leaf} illustrates the two-dimensional footprints of leaves (integrated from their three-dimensional structures) overlaid on the integrated intensity map of \nh.
The NW subcluster has a similar number in high-contrast and low-contrast leaves, while the number of high-contrast leaves is noticeably less than that of the low-contrast leaves in the SE subcluster.
The sprouts are mostly distributed in the outskirts of the two subclusters.

The northwest corner in Figure~\ref{fig:serpens_n2hp} showing weak \nh \ emission is associated with a high-contrast leaf (leaf 5), a low-contrast leaf (leaf 19), and a 3~mm source (S1).
Dust emission is detected in this region at 250 \um, 350 \um, and 500 \um \ with \textit{Herschel} (see Fig.~\ref{fig:pointing} for the 250 \um \ image). 
\citet{2000ApJ...536..845M} also showed a C$^{18}$O(1-0) peak near the \nh \ peak.
This is likely a new location for star formation.

The broad outline of the branching structure and the complexity of the tree can be captured in a few statistical measures \citep{HS92}. 
A branching level is defined as the number of branching steps between the object and the tree base.  
For example, leaf 3 has a branching level of zero because it grows directly from the tree base.
Leaf 36 has a branching level of two since it goes through branch 59 and 62 before reaching the base branch.    
The branching levels for each structure are summarized in Table~\ref{tbl:tree}. 
The maximum branching level is eight in the SE subcluster, and five in the NW subcluster.
The mean branching level of the entire tree is 3.1, while the mean branching level of the SE and NW subcluster is 5.0 and 3.7, respectively.

A path length is defined as the number of branching steps between a leaf and the tree base, 
similar to the definition of a branching level.  
Different from a branching level, a path length is considered only for leaves and not considered for branches. 
The mean path length of the tree, defined as the mean of path lengths from the leaves, 
can then better reflect the hierarchy in the tree since branching levels from branches would not be double counted.
Larger mean path length corresponds a larger degree of hierarchical structure in the tree.
The mean path length of the entire tree is 3.0, while the mean path length of the SE and NW subcluster is 5.1 and 4.1, respectively.
This suggests that the SE subcluster is more hierarchical than the NW subcluster.  

The branching ratio is defined as the number of substructures which merge at a branch level.
For example, branch 55 has a branching ratio of three since it joins leaf 6, 7, and 12. Branch 60 has a branching ratio of two since it fragments to leaf 40 and branch 49. A larger branching ratio corresponds to a higher degree of fragmentation.
The mean branching ratio in Serpens Main is 3.2 (2.6~in the SE subcluster and 2.4~in the NW subcluster), smaller than the mean branching ratio of 3.9~in Barnard 1 (Paper I).

Overall, these tree statistics indicate that the SE subcluster exhibits more complex hierarchical structure compared to the NW subcluster.
There are $\sim40$ YSOs in the SE subcluster and $\sim12$ YSOs in the NW subcluster \citep{2009ApJS..181..321E}, suggesting that the complexity of hierarchical structure is associated with star formation activity.  
The Serpens Main dendrogram also has more complex structure compared to the Barnard 1 dendrogram presented in Paper I; 
that dendrogram has a maximum branching level of 4, a mean path length of 1.2, and a mean branching ratio of 3.9. 
Since Serpens Main has more star formation activity than Barnard 1, these results suggest that the hierarchical nature of the dense gas in molecular clouds is linked with the star formation activity of those regions. 

The comparison in the distribution between the leaves and 3~mm sources (Sect.~\ref{sect:conti}) is shown in Figure \ref{fig:leaf}.
The 3~mm continuum sources are indicated by star symbols (including yellow and red stars); the red stars represent the sources coincident with the SMM cores.
A majority of the SMM cores are better associated with the high-contrast leaves than the low-contrast leaves, suggesting that the high-contrast leaves may be associated with formation of dense cores.

\subsection{Morphological Properties}
\label{sect:dendromor}

Table~\ref{tbl:tree} shows the morphological properties of the leaves and branches.
We derived the morphological properties using the two-dimensional footprint (as shown in Fig.~\ref{fig:leaf}) of inherently three-dimensional structures.
RA, Dec, major axis, minor axis, and position angle were calculated with the task ``{\tt regionprops}'' in MATLAB.
For all the objects including leaves and branches, 
we include all the emission from the leaves/branches within it when performing the regionprops fitting.
Axis ratios are minor axes divided by major axes, and the sizes are the geometric mean of the two axes.
We define a ``filling factor" to quantify the regularity of an object compared to its fitted shape.
The filling factor is calculated as the area of the object enclosed by the fitted ellipse divided by the area of the fitted ellipse; smaller values correspond to more irregular shapes.

Figure~\ref{fig:histo} presents the histograms of the size, filling factor, and axis ratio of the leaves and branches.
The leaves have sizes ranging from 0.024 pc ($\sim 1.7$ beam size) to 0.105 pc with a mean of 0.053 pc from all the leaves, and the branches have larger sizes with a mean value of 0.26 pc.
The high-contrast leaves have a moderately larger mean size of 0.071 pc than the low-contrast leaves with a mean size of 0.046 pc.  
The histogram of filling factor for leaves displays an increase in number toward regular shapes (value close to 1); branches show an opposite trend.
The flat distribution in axis ratio shows that the objects are rarely described by spherical shapes (axis ratio = 1), and a number of objects have elongated structures.
In particular, the southern filament shows several high-contrast and low-contrast leaves with their branches in filamentary morphologies (Fig.~\ref{fig:leaf}).

\subsection{Kinematic Properties}
\label{sect:dendrokin}

\begin{figure}[t!]
\begin{center}
\includegraphics[scale=2.4]{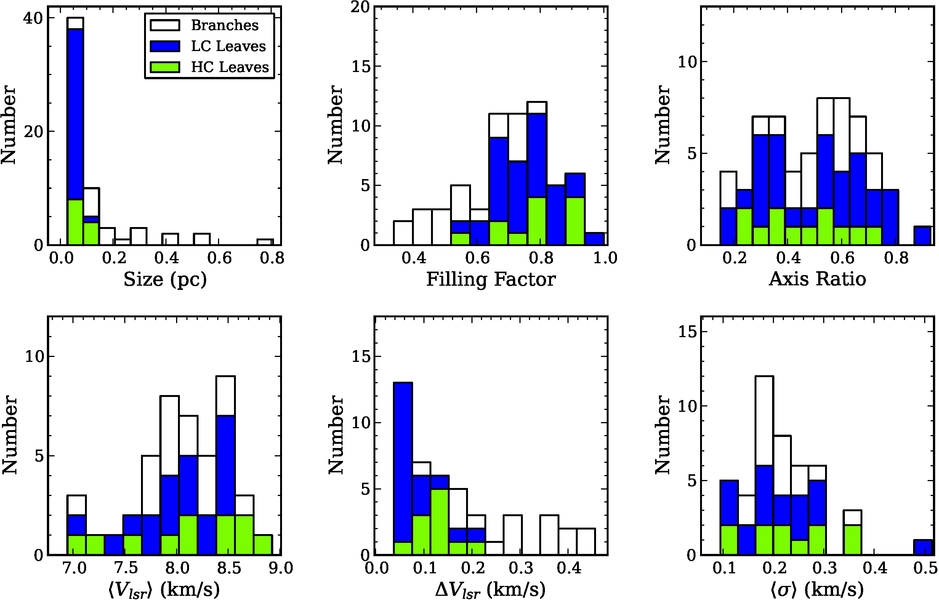}
\caption{\footnotesize Histograms of leaf and branch properties: size, filling parameter, axis ratio, mean $\text{V}_{\text{lsr}}$ ($\langle \text{V}_{\text{lsr}} \rangle$), $\text{V}_{\text{lsr}}$ variation ($\Delta \text{V}_{\text{lsr}}$), and velocity dispersion ($\langle \sigma \rangle$).  We separate high-contrast (HC) leaves (green), low-contrast (LC) leaves (blue; including the sprouts), and branches (white).}
\label{fig:histo}
\end{center}
\end{figure}

Table~\ref{tbl:tree} and Figure~\ref{fig:histo} also show the kinematic properties of the leaves and branches including $\langle \text{V}_{\text{lsr}} \rangle$, $\Delta \text{V}_{\text{lsr}}$, $\langle \sigma \rangle$, and $\Delta \sigma$.
All of these quantities are derived using the results of the \nh \ spectral line fitting described in Sect.~\ref{sect:kinematics}.
$\langle \text{V}_{\text{lsr}} \rangle$ and $\langle \sigma \rangle$ are the mean centroid velocities ($\text{V}_{\text{lsr}}$) and the mean velocity dispersion ($\sigma$) in an object; the calculated means are weighted by the statistical uncertainties from the spectral line fitting.
$\Delta \text{V}_{\text{lsr}}$, the $\text{V}_{\text{lsr}}$ variation, is the standard deviation of $\text{V}_{\text{lsr}}$ to the mean value ($\langle \text{V}_{\text{lsr}} \rangle$).
Similarly, $\Delta \sigma$ is the standard deviation of velocity dispersions to the mean value ($\langle \sigma \rangle$).

In determining the object kinematic properties,
the calculation was based on the spatial footprint of a leaf; for a branch, it was based on the area in the ``onion layer", which is the area between the branch and the structures directly above it in the tree.
The regions with two-velocity components (Sect.~\ref{sect:kinematics}) were not used in this analysis due to the overlapping spatial footprints.
We also excluded objects that had less than three beam areas of kinematic pixels.

The mean $\text{V}_{\text{lsr}}$ ($\langle \text{V}_{\text{lsr}} \rangle$) ranges from 6.9 to 8.8 km~s$^{-1}$ without a significant difference in distribution between leaves and branches (Fig.~\ref{fig:histo}).
Leaves and branches in the NW subcluster have $\langle \text{V}_{\text{lsr}} \rangle$ distributed between 8.0 to 8.5 km~s$^{-1}$, and the objects in the SE subcluster are more blue-shifted except for the southern filaments.
In the $\Delta \text{V}_{\text{lsr}}$ distribution ($\text{V}_{\text{lsr}}$ variation calculated as the standard deviation of $\text{V}_{\text{lsr}}$ inside an object), the branches clearly show larger $\Delta \text{V}_{\text{lsr}}$ than the leaves.
Branches and leaves have a mean $\Delta \text{V}_{\text{lsr}}$ of 0.28 km~s$^{-1}$ and 0.09 km~s$^{-1}$, respectively.
This is expected since branches incorporate larger spatial scales and may reflect large-scale motions.
The high-contrast leaves have larger $\Delta \text{V}_{\text{lsr}}$ peaking at $\sim 0.13$ km~s$^{-1}$ than the low-contrast leaves peaking at $\sim0.06$ km~s$^{-1}$.
Similarly, this difference is possibly due to larger sizes for the high-contrast leaves which incorporate more turbulent power across the plane of the sky (see Sect.~\ref{sect:denkin}), and/or due to local star formation activities.

For $\langle \sigma \rangle$, there is no clear difference in the distribution between leaves and branches, and there is also no clear difference in the distribution between the high-contrast and low-contrast leaves.
Most of the leaves and branches show velocity dispersions below the sonic level, 0.27 km~s$^{-1}$ assuming 20 K gas temperature.
The mean velocity dispersion including leaves and branches is 0.2 km~s$^{-1}$.
A few high-contrast leaves have supersonic velocity dispersions (leaf 9, 20, 33, 44).
Among these leaves, leaf 9, 33, and 44 have the highest intensities in the tree.
Leaf 20, 33, 44 are located in the central region of the SE subcluster and are associated with SMM cores; leaf 9 is associated with SMM1 in the NW subcluster.
The close correlation between the four leaves and the SMM cores suggest that the large velocity dispersions may be due to local star formation activity.
Most of the leaves have $\langle \sigma \rangle$ 2 to 3 times larger than $\Delta \text{V}_{\text{lsr}}$.

\begin{deluxetable}{lllccccccccccccc}
\tabletypesize{\tiny}
\setlength{\tabcolsep}{0.03in}
\tablecaption{\nh \ Dendrogram Leaf and Branch Properties}
\tablewidth{0pt}
\tablecolumns{16}
\tablehead{
\colhead{No.} & \colhead{RA\tablenotemark{a}} & \colhead{Dec\tablenotemark{a}} & \colhead{Maj.\ Axis\tablenotemark{a}} & \colhead{Min.\ Axis\tablenotemark{a}} & \colhead{PA\tablenotemark{a}} & \colhead{Axis\tablenotemark{b}} & \colhead{Filling\tablenotemark{c}} & \colhead{Size\tablenotemark{d}} & \colhead{$\langle \text{V}_{\text{lsr}} \rangle$\tablenotemark{e}} & \colhead{$\Delta \text{V}_{\text{lsr}}$\tablenotemark{f}} & \colhead{$\langle \sigma \rangle$\tablenotemark{g}} & \colhead{$\Delta \sigma$\tablenotemark{h}} & \colhead{Pk.\ Int.\ \tablenotemark{i}} & \colhead{Contrast\tablenotemark{j}} & \colhead{Level\tablenotemark{k}} \\
\colhead{} & \colhead{(J2000)} & \colhead{(J2000)} & \colhead{($\arcsec$)} & \colhead{($\arcsec$)} & \colhead{($\deg$)} & \colhead{Ratio} & \colhead{Factor} & \colhead{(pc)} & \colhead{(km~s$^{\text{-1}}$)} & \colhead{(km~s$^{\text{-1}}$)} & \colhead{(km~s$^{\text{-1}}$)} & \colhead{(km~s$^{\text{-1}}$)} & \colhead{(Jy~beam$^{\text{-1}}$)} & \colhead{($\sigma$)} & \colhead{}
}
\startdata
\multicolumn{15}{c}{\bf Leaves} \\
\hline
0 & 18:30:03.2 & +01:14:36.5 & 37.3 & 14.7 & 146.9 &  0.39 &  0.76 & 0.047 & \nodata & \nodata & \nodata & \nodata &  0.56 &  2.7 & 0	\\
1 & 18:29:44.6 & +01:17:04.4 & 29.6 & 25.8 & 61.3 &  0.87 &  0.61 & 0.056 & \nodata & \nodata & \nodata & \nodata &  0.58 &  2.8 & 0	\\
2 & 18:29:57.0 & +01:13:09.4 & 20.5 & 7.8 & 40.4 &  0.38 &  0.86 & 0.025 & \nodata & \nodata & \nodata & \nodata &  1.54 &  3.8 & 5	\\
3 & 18:30:02.0 & +01:11:37.1 & 32.1 & 9.2 & 74.8 &  0.29 &  0.74 & 0.035 & \nodata & \nodata & \nodata & \nodata &  0.51 &  2.2 & 0	\\
4 & 18:29:43.7 & +01:16:42.5 & 40.0 & 21.4 & 61.6 &  0.53 &  0.60 & 0.059 & \nodata & \nodata & \nodata & \nodata &  0.54 &  2.4 & 0	\\
5 & 18:29:30.8 & +01:18:58.4 & 70.5 & 38.4 & 148.5 &  0.54 &  0.79 & 0.105 &  8.83(2) &  0.12(1) &  0.12(0) &  0.03(0) &  1.95 & 14.0 & 1	\\
6 & 18:29:59.3 & +01:10:09.8 & 77.5 & 18.0 & 10.0 &  0.23 &  0.67 & 0.075 &  8.61(0) &  0.04(0) &  0.11(0) &  0.02(0) &  1.48 &  6.2 & 4	\\
7 & 18:30:00.4 & +01:11:26.7 & 76.9 & 11.6 &  0.7 &  0.15 &  0.79 & 0.060 &  8.48(1) &  0.05(1) &  0.19(1) &  0.03(0) &  1.27 &  4.3 & 4	\\
8 & 18:30:04.8 & +01:14:49.8 & 29.0 & 16.7 & 35.2 &  0.58 &  0.68 & 0.044 &  7.95(9) &  0.21(8) &  0.28(7) &  0.16(6) &  0.60 &  2.2 & 1	\\
9 & 18:29:49.6 & +01:15:18.3 & 39.0 & 22.2 & 145.3 &  0.57 &  0.75 & 0.059 &  8.52(3) &  0.11(2) &  0.35(3) &  0.12(2) &  2.85 & 12.8 & 4	\\
10 & 18:29:52.0 & +01:15:59.8 & 42.5 & 9.1 & 11.1 &  0.22 &  0.73 & 0.040 &  8.52(1) &  0.04(1) &  0.19(0) &  0.02(0) &  1.71 &  3.3 & 5	\\
11 & 18:30:02.1 & +01:08:44.0 & 94.7 & 28.3 & 28.7 &  0.30 &  0.74 & 0.104 &  8.44(1) &  0.04(0) &  0.10(0) &  0.02(0) &  0.95 &  5.3 & 1	\\
12 & 18:30:00.6 & +01:10:21.2 & 41.2 & 14.5 &  3.6 &  0.35 &  0.85 & 0.049 &  8.36(1) &  0.04(1) &  0.11(0) &  0.02(0) &  1.44 &  5.8 & 4	\\
13 & 18:29:48.7 & +01:14:31.9 & 41.4 & 29.3 & 133.4 &  0.71 &  0.56 & 0.070 &  8.41(8) &  0.17(7) &  0.48(7) &  0.16(6) &  0.62 &  3.2 & 0	\\
14 & 18:29:49.4 & +01:15:40.2 & 23.6 & 18.0 & 105.5 &  0.76 &  0.77 & 0.041 &  8.17(3) &  0.06(2) &  0.25(1) &  0.03(1) &  1.82 &  3.3 & 4	\\
15 & 18:29:44.5 & +01:16:00.7 & 28.6 & 15.8 & 95.8 &  0.55 &  0.74 & 0.043 &  8.42(3) &  0.07(2) &  0.28(1) &  0.04(1) &  0.71 &  3.2 & 1	\\
16 & 18:29:50.6 & +01:16:50.0 & 84.4 & 22.6 & 27.7 &  0.27 &  0.58 & 0.088 &  8.53(2) &  0.12(1) &  0.21(1) &  0.05(0) &  2.05 &  6.5 & 5	\\
17 & 18:29:48.4 & +01:16:39.3 & 42.7 & 30.3 & 128.5 &  0.71 &  0.76 & 0.072 &  8.58(3) &  0.15(2) &  0.30(2) &  0.13(2) &  2.22 &  9.0 & 4	\\
18 & 18:29:51.5 & +01:17:42.8 & 47.5 & 18.1 & 29.9 &  0.38 &  0.64 & 0.059 & \nodata & \nodata & \nodata & \nodata &  0.54 &  2.5 & 0	\\
19 & 18:29:28.4 & +01:18:31.0 & 26.4 & 18.2 & 59.7 &  0.69 &  0.68 & 0.044 & \nodata & \nodata & \nodata & \nodata &  0.79 &  3.4 & 1	\\
20 & 18:29:56.4 & +01:13:08.2 & 26.9 & 12.3 & 76.9 &  0.46 &  0.89 & 0.037 &  8.12(4) &  0.10(3) &  0.37(4) &  0.09(3) &  1.99 &  7.9 & 5	\\
21 & 18:30:00.1 & +01:13:10.2 & 13.3 & 10.8 & 40.9 &  0.81 &  0.88 & 0.024 & \nodata & \nodata & \nodata & \nodata &  0.87 &  2.6 & 3	\\
22 & 18:30:09.3 & +01:13:23.8 & 22.2 & 15.0 & 39.1 &  0.68 &  0.78 & 0.037 & \nodata & \nodata & \nodata & \nodata &  0.82 &  4.1 & 1	\\
23 & 18:29:56.7 & +01:13:41.3 & 22.5 & 8.2 & 24.6 &  0.36 &  0.82 & 0.027 &  8.00(2) &  0.04(2) &  0.22(2) &  0.05(2) &  1.51 &  2.5 & 8	\\
24 & 18:29:48.6 & +01:13:38.4 & 24.0 & 16.5 & 125.6 &  0.69 &  0.85 & 0.040 & \nodata & \nodata & \nodata & \nodata &  0.53 &  2.3 & 0	\\
25 & 18:29:53.3 & +01:13:59.9 & 79.3 & 29.3 & 77.8 &  0.37 &  0.65 & 0.097 &  8.13(1) &  0.10(1) &  0.18(0) &  0.03(0) &  1.48 &  6.2 & 5	\\
26 & 18:29:52.2 & +01:15:24.3 & 45.6 & 20.5 & 43.3 &  0.45 &  0.89 & 0.061 &  7.99(3) &  0.12(2) &  0.20(1) &  0.06(1) &  2.55 & 12.0 & 4	\\
27 & 18:29:57.1 & +01:15:18.5 & 35.6 & 18.2 & 74.7 &  0.51 &  0.75 & 0.051 &  8.31(4) &  0.13(3) &  0.18(1) &  0.05(1) &  1.10 &  5.7 & 2	\\
28 & 18:29:59.7 & +01:15:20.9 & 57.6 & 10.3 & 135.3 &  0.18 &  0.65 & 0.049 &  7.98(4) &  0.11(3) &  0.21(1) &  0.04(1) &  1.44 &  4.8 & 6	\\
29 & 18:29:49.2 & +01:16:14.3 & 26.5 & 20.0 & 135.4 &  0.76 &  0.73 & 0.046 &  8.18(2) &  0.05(1) &  0.20(1) &  0.04(1) &  1.65 &  4.8 & 3	\\
30 & 18:29:50.5 & +01:18:24.3 & 42.3 & 20.5 & 113.3 &  0.48 &  0.67 & 0.059 &  8.03(3) &  0.06(3) &  0.13(2) &  0.05(2) &  0.66 &  3.5 & 0	\\
31 & 18:29:56.0 & +01:12:22.5 & 25.7 & 19.1 & 53.9 &  0.74 &  0.64 & 0.045 & \nodata & \nodata & \nodata & \nodata &  0.51 &  2.2 & 0	\\
32 & 18:30:13.6 & +01:13:46.4 & 55.1 & 16.1 & 143.6 &  0.29 &  0.68 & 0.060 & \nodata & \nodata & \nodata & \nodata &  0.90 &  5.8 & 0	\\
33 & 18:29:58.6 & +01:14:03.9 & 67.3 & 39.9 & 170.5 &  0.59 &  0.80 & 0.104 &  7.66(2) &  0.13(1) &  0.26(0) &  0.06(0) &  3.86 & 24.0 & 8	\\
34 & 18:30:11.3 & +01:15:59.3 & 34.6 & 21.1 & 131.0 &  0.61 &  0.87 & 0.054 &  7.83(2) &  0.06(1) &  0.09(0) &  0.01(0) &  0.89 &  5.0 & 1	\\
35 & 18:30:13.2 & +01:16:12.5 & 21.0 & 12.2 & 155.5 &  0.58 &  0.80 & 0.032 & \nodata & \nodata & \nodata & \nodata &  0.63 &  2.6 & 1	\\
36 & 18:30:03.0 & +01:11:41.4 & 18.1 & 12.3 & 94.7 &  0.68 &  0.91 & 0.030 & \nodata & \nodata & \nodata & \nodata &  0.91 &  4.0 & 2	\\
37 & 18:30:02.5 & +01:12:15.6 & 22.6 & 8.7 & 99.0 &  0.39 &  0.81 & 0.028 & \nodata & \nodata & \nodata & \nodata &  0.96 &  3.4 & 3	\\
38 & 18:30:01.3 & +01:12:08.3 & 34.6 & 10.3 &  2.5 &  0.30 &  0.85 & 0.038 &  7.32(2) &  0.05(2) &  0.15(2) &  0.04(1) &  1.27 &  3.3 & 6	\\
39 & 18:29:49.8 & +01:14:17.3 & 15.1 & 11.9 & 67.6 &  0.78 &  0.95 & 0.027 &  7.50(3) &  0.06(2) &  0.30(4) &  0.09(4) &  0.76 &  2.6 & 2	\\
40 & 18:29:47.2 & +01:14:48.0 & 26.9 & 8.3 & 38.4 &  0.31 &  0.79 & 0.030 &  7.71(5) &  0.10(4) &  0.24(6) &  0.11(5) &  0.71 &  2.1 & 2	\\
41 & 18:30:00.1 & +01:11:34.3 & 46.7 & 13.3 & 125.5 &  0.29 &  0.89 & 0.050 & \nodata & \nodata & \nodata & \nodata &  2.11 & 11.0 & 6	\\
42 & 18:29:59.5 & +01:12:43.2 & 23.3 & 12.6 & 14.7 &  0.54 &  0.93 & 0.034 &  7.02(3) &  0.09(3) &  0.26(2) &  0.05(2) &  1.74 &  5.6 & 8	\\
43 & 18:29:58.9 & +01:13:20.8 & 31.8 & 11.4 &  3.1 &  0.36 &  0.91 & 0.038 &  7.25(5) &  0.12(5) &  0.23(3) &  0.07(2) &  2.18 &  8.7 & 8	\\
44 & 18:30:00.7 & +01:12:59.9 & 40.0 & 27.4 &  7.3 &  0.69 &  0.78 & 0.067 &  6.95(5) &  0.22(4) &  0.28(2) &  0.08(1) &  3.11 & 18.2 & 8	\\
\hline	
\multicolumn{15}{c}{\bf Branches} \\
\hline
45 & 18:29:49.5 & +01:15:25.6 & 58.7 & 34.8 &  7.3 &  0.59 &  0.78 & 0.091 &  8.31(4) &  0.16(3) &  0.28(2) &  0.10(1) &  1.46 &  \nodata & 3	\\
46 & 18:29:50.9 & +01:16:38.9 & 124.2 & 24.4 & 25.2 &  0.20 &  0.55 & 0.111 &  8.59(5) &  0.17(4) &  0.19(1) &  0.05(1) &  1.35 &  \nodata & 4	\\
47 & 18:29:50.4 & +01:16:24.3 & 165.2 & 65.0 & 28.1 &  0.39 &  0.51 & 0.208 &  8.52(6) &  0.23(4) &  0.23(2) &  0.09(1) &  1.24 &  \nodata & 3	\\
48 & 18:29:58.5 & +01:13:59.5 & 103.9 & 57.2 & 179.7 &  0.55 &  0.52 & 0.155 &  8.00(6) &  0.29(5) &  0.25(1) &  0.08(1) &  1.24 &  \nodata & 7	\\
49 & 18:29:50.1 & +01:16:08.1 & 176.9 & 110.6 & 19.3 &  0.63 &  0.75 & 0.281 &  8.34(2) &  0.28(1) &  0.21(0) &  0.11(0) &  1.13 &  \nodata & 2	\\
50 & 18:30:00.5 & +01:12:56.3 & 53.4 & 35.8 & 146.0 &  0.67 &  0.74 & 0.088 &  7.05(8) &  0.19(7) &  0.25(5) &  0.12(4) &  1.13 &  \nodata & 7	\\
51 & 18:29:56.7 & +01:13:09.4 & 49.9 & 31.2 & 78.0 &  0.62 &  0.74 & 0.079 &  8.27(5) &  0.20(4) &  0.36(4) &  0.19(3) &  1.13 &  \nodata & 4	\\
52 & 18:29:59.2 & +01:13:39.6 & 156.2 & 66.3 & 19.6 &  0.42 &  0.54 & 0.205 &  7.84(8) &  0.37(6) &  0.21(2) &  0.10(1) &  1.02 &  \nodata & 6	\\
53 & 18:30:00.6 & +01:11:47.5 & 85.6 & 24.4 & 149.3 &  0.29 &  0.57 & 0.092 & \nodata & \nodata & \nodata & \nodata &  0.91 &  \nodata & 5	\\
54 & 18:29:59.4 & +01:13:53.7 & 201.0 & 93.2 &  0.9 &  0.46 &  0.43 & 0.275 &  7.91(6) &  0.34(4) &  0.23(1) &  0.08(1) &  0.91 &  \nodata & 5	\\
55 & 18:30:00.0 & +01:10:48.0 & 193.6 & 36.3 & 175.3 &  0.19 &  0.60 & 0.169 &  8.49(2) &  0.10(1) &  0.14(1) &  0.07(1) &  0.80 &  \nodata & 3	\\
56 & 18:29:58.6 & +01:13:43.4 & 242.4 & 163.2 & 12.3 &  0.67 &  0.34 & 0.400 &  7.92(6) &  0.37(4) &  0.20(1) &  0.08(1) &  0.80 &  \nodata & 4	\\
57 & 18:29:58.7 & +01:13:40.3 & 233.2 & 173.3 & 10.8 &  0.74 &  0.41 & 0.404 &  7.82(4) &  0.39(3) &  0.20(1) &  0.09(0) &  0.69 &  \nodata & 3	\\
58 & 18:29:59.2 & +01:13:10.1 & 351.4 & 185.1 &  5.0 &  0.53 &  0.39 & 0.513 &  7.90(4) &  0.43(3) &  0.20(1) &  0.11(0) &  0.58 &  \nodata & 2	\\
59 & 18:29:59.4 & +01:13:14.9 & 359.7 & 207.0 &  5.1 &  0.58 &  0.44 & 0.549 &  7.83(4) &  0.44(3) &  0.19(1) &  0.11(0) &  0.47 &  \nodata & 1	\\
60 & 18:29:50.2 & +01:16:05.0 & 180.1 & 129.2 & 16.3 &  0.72 &  0.75 & 0.307 &  8.16(4) &  0.27(3) &  0.16(1) &  0.09(1) &  0.47 &  \nodata & 1	\\
61 & 18:29:30.3 & +01:18:55.2 & 100.6 & 50.2 & 138.6 &  0.50 &  0.65 & 0.143 & \nodata & \nodata & \nodata & \nodata &  0.42 &  \nodata & 0	\\
62 & 18:29:56.9 & +01:14:02.0 & 553.7 & 260.4 & 28.4 &  0.47 &  0.47 & 0.764 &  8.07(4) &  0.42(3) &  0.20(1) &  0.11(0) &  0.37 &  \nodata & 0	\\
63 & 18:30:11.8 & +01:16:02.5 & 61.9 & 21.8 & 118.5 &  0.35 &  0.68 & 0.074 & \nodata & \nodata & \nodata & \nodata &  0.35 &  \nodata & 0	
\enddata
\label{tbl:tree}
\tablenotetext{a}{The coordinate (RA and Dec), major axis, minor axis, and position angle are derived from the task \textit{regionprops} in \textit{MATLAB}.}
\tablenotetext{b}{The axis ratio is calculated as minor axis divided by major axis.}
\tablenotetext{c}{The filling factor is calculated as the ratio between the area of the object enclosed by the fitted ellipse and the area of the fitted ellipse.}
\tablenotetext{d}{The size is the geometric mean of the major and minor axis, calculated as $\sqrt{\text{(Major \ axis)} \times \text{(Minor \ axis)}}$.}
\tablenotetext{e}{The mean $\text{V}_{\text{lsr}}$ ($\langle \text{V}_{\mathrm{lsr}} \rangle$) is the mean of fitted $\text{V}_{\text{lsr}}$ in an object weighted by the statistical uncertainty from $\text{V}_{\text{lsr}}$ fitting.  The uncertainty in the weighted mean $\text{V}_{\text{lsr}}$ (shown in parentheses) is computed as $\Delta \text{V}_{\text{lsr}} / \sqrt{N}$, where $N$ is the number of independent beams.  The uncertainty is reported in the last digit.}
\tablenotetext{f}{The $\text{V}_{\text{lsr}}$ variation ($\Delta \text{V}_{\text{lsr}}$) is calculated as the standard deviation of $\text{V}_{\text{lsr}}$ in an object.
The uncertainty in $\Delta \text{V}_{\text{lsr}}$ (shown in parentheses) is calculated as $\Delta \text{V}_{\text{lsr}} / \sqrt{2(N-1)}$.  The uncertainty is reported in the last digit.}
\tablenotetext{g}{The mean velocity dispersion ($\langle \sigma \rangle$) is calculated as the mean of fitted velocity dispersions in an object weighted by the statistical uncertainty from $\sigma$ fitting.  The uncertainty in the weighted mean $\sigma$ (shown in parentheses) is computed as $\Delta \sigma / \sqrt{N}$, where $N$ is the number of independent beams.  The uncertainty is reported in the last digit.}
\tablenotetext{h}{The $\sigma$ variation ($\Delta \sigma$) is calculated as the standard deviation of $\sigma$ in an object.
The error in $\sigma$ (shown in parentheses) is calculated as $\Delta \sigma / \sqrt{2(N-1)}$.  The error is reported to the last digit.}
\tablenotetext{i}{For a leaf, the peak intensity in a single channel from our binned data cube. For a branch, the intensity level where the structures directly above it merge together.}
\tablenotetext{j}{The contrast is calculated as the difference between the peak intensity and the merging level of an object.  Contrasts are only computed for leaves, divided by the 1$\sigma$ sensitivity of the data.}
\tablenotetext{k}{The branching level in the dendrogram. For example, level 0 means that the object grows from the base level; level 1 means that the object grows from a branch one level above the base level.}
\end{deluxetable}
\clearpage

\section{Implications from Linewidth-Size Relations: The Characteristic Cloud Depth}
\label{sect:denkin}

From the dendrogram analysis (Sect.~\ref{sect:dendro}), we obtained two important measures of the gas kinematic motions: the mean velocity dispersion in the object ($\langle \sigma \rangle$) and the $\text{V}_{\text{lsr}}$ variation across the object ($\Delta \text{V}_{\text{lsr}}$), for each leaf and branch.
The velocity dispersion ($\langle \sigma \rangle$) is calculated as the mean velocity dispersion for each pixel of the object on the plane of the sky; it represents the mean gas motions along the line of sight of each object.
The $\text{V}_{\text{lsr}}$ variation ($\Delta \text{V}_{\text{lsr}}$) is calculated as the standard deviation of the mean $\text{V}_{\text{lsr}}$ inside an object (leaf or branch); it represents the gas motions across different spatial scales on the plane of the sky.
Traditionally,  the ``linewidth-size'' relation is summarized as ``Larson's Law'' \citep{1981MNRAS.194..809L}, which expresses a power-law dependence between velocity linewidths and sizes \citep[e.g.,][]{1983ApJ...266..309M,1987ApJ...319..730S,1992ApJ...384..523F}.
We denote the linewidths in Larson's law as ``undecomposed linewidths'' since the linewidths were measured with low angular and spatial resolutions in the foundation studies.
With our angular resolution of 7\arcsec \ (spatial resolution of 2900 AU),
we are able to dissect undecomposed linewidths into $\langle \sigma \rangle$ and $\Delta \text{V}_{\text{lsr}}$, which reflect gas motions along the line of sight and variation across the plane of the sky.

\begin{figure}
\begin{center}
\includegraphics[scale=2.1]{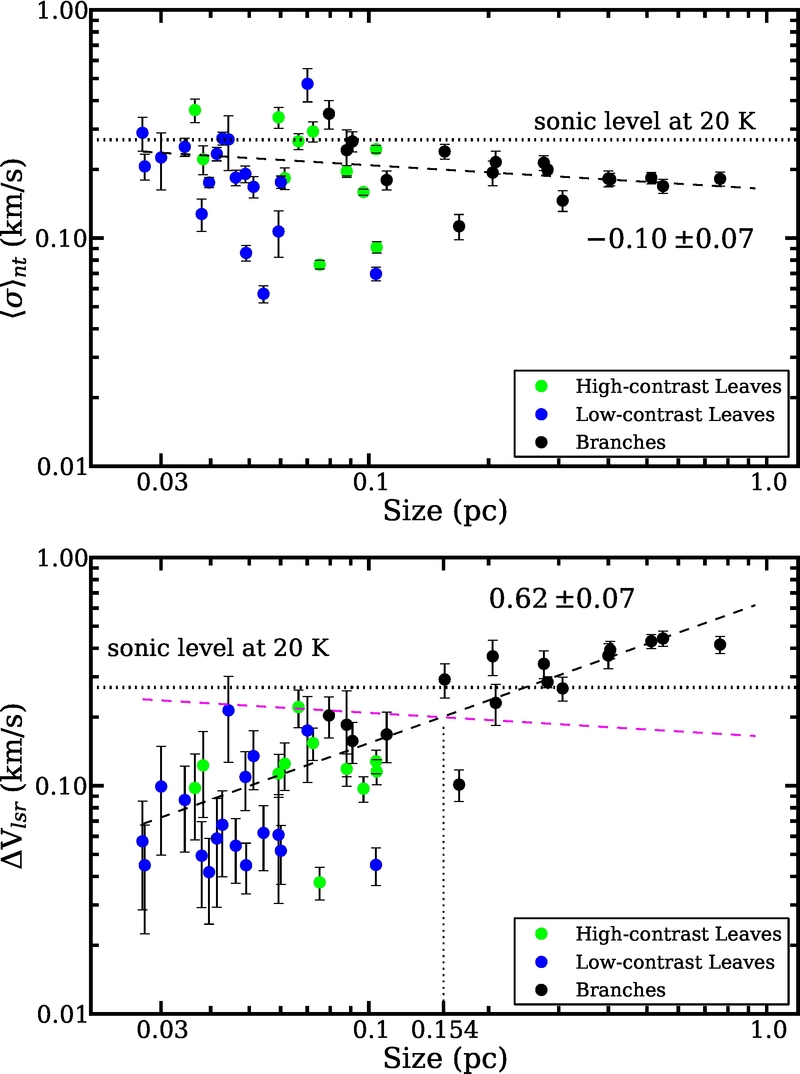}
\caption{\footnotesize
\textit{Top panel:} The nonthermal velocity dispersion vs.\ size relation including the leaves and branches based on the data in Table 4.  The data nearly shows no correlation (power-law index $=-0.10 \pm 0.07$). \textit{Bottom panel:} The $\text{V}_{\text{lsr}}$ variation vs.\ size relation including the leaves and branches.  A positive correlation is observed.  The best-fitted power-law index is $0.62 \pm 0.07$.  The magenta dashed line shows the best-fit of the nonthermal velocity dispersion vs.\ size relation (top panel).  The two power-law relations intersect at a size scale of 0.15~pc.
}
\label{fig:denkin}
\end{center}
\end{figure}

Figure~\ref{fig:denkin} shows the two relations: $\langle \sigma \rangle_{\text{nt}}$ vs.\ size and $\Delta \text{V}_{\text{lsr}}$ vs.\ size, where $\langle \sigma \rangle_{\text{nt}}$ is the nonthermal component of the velocity dispersion and the size is the geometric mean of major and minor axes for each object, based on the data in Table~\ref{tbl:tree}.
We have included both leaves and branches in Fig.~\ref{fig:denkin} to study the statistical behavior of kinematics across the full range of spatial scales in our data.
The nonthermal velocity dispersion ($\langle \sigma \rangle _{\text{nt}}$) is calculated as $\sqrt{\langle \sigma \rangle^{2}-kT/(\mu m_{H})}$ by assuming 20 K \citep{2000ApJ...536..845M} for the temperature T, where $k$ is the Boltzmann constant, $\mu=29.0$ is the mean molecular weight for \nh, and $m_{H}$ is the hydrogen mass.
The sonic velocity dispersion at 20 K is 0.27 km~s$^{-1}$ as shown in the dotted line.
The majority of the objects have nonthermal velocity dispersions below the sonic level yet well above the \nh \ thermal velocity dispersion of 0.075 km~s$^{-1}$.
This shows that the gas motions are subsonic to sonic along the line of sight.

The best-fit, assuming a power-law relation in $\langle \sigma \rangle_{\text{nt}}$ vs.\ size, shows a power-law index of $-0.07\pm 0.07$ (top panel in Fig.~\ref{fig:denkin}).
In other words, the nonthermal velocity dispersion is roughly constant with size.
On the other hand, 
the $\text{V}_{\text{lsr}}$ variations ($\Delta \text{V}_{\text{lsr}}$) shows a clear power-law dependence with size.
The best-fit power-law index is $0.61\pm0.08$ (bottom panel in Fig.~\ref{fig:denkin}).
The power-law relation is expected since $\Delta \text{V}_{\text{lsr}}$ captures the gas motions across different spatial scales on the plane of the sky.

The difference between these two relations can be used to gain insight about the cloud structure.
Assuming that the gas motions are characterized by isotropic three dimensional turbulence, the velocity dispersion and the V$_{\text{lsr}}$ variation across the sky result from the observational manifestation of the turbulent power spectrum (see Paper I for a detailed discussion). Since the turbulent power increases to larger spatial scales, the observed velocity dispersion has a correspondence with the largest spatial scale sampled by the observation. 
For example, the velocity dispersion along a line of sight ($\langle \sigma \rangle_{\text{nt}}$), is characteristic of the larger of the beam linear size and the depth of the emission region. If an object has larger $\langle \sigma \rangle_{\text{nt}}$ compared to $\Delta \text{V}_{\text{lsr}}$, it suggests that the spatial scale along the line of sight exceeds the spatial scale sampled on the plane of the sky.
On the other hand, if an object has smaller $\langle \sigma \rangle_{\text{nt}}$ compared to $\Delta \text{V}_{\text{lsr}}$, the structure is more extended on the plane of the sky compared to its depth. 
Statistically, the intersection of the two power-laws corresponds to an approximately equal spatial scale in both dimensions and therefore provides an estimate on the depth.

Since velocity dispersions increasing with spatial scales is a characteristics of turbulent power, the lack of a nonthermal velocity dispersion vs.\ size relation suggests that statistically the objects have similar spatial scales into the plane of the sky, i.e., similar depths into the sky.
Figure~\ref{fig:denkin} (bottom plot) shows that the two power-laws (black and magenta dashed lines) intersect at a size scale of 0.15~pc, suggesting that the objects have a characteristic depth of 0.15~pc.
Statistically, the objects with projected sizes smaller than this characteristic scale (mostly leaves) have smaller projected sizes on the plane of the sky compared to their depths into the sky since they have smaller $\Delta \text{V}_{\text{lsr}}$ compared to $\langle \sigma \rangle_{\text{nt}}$.  
On the other hand, the objects with projected sizes larger than this characteristic scale (branches) have larger projected sizes on the plane of the sky compared to their depths into the sky.
Similar power-law indices in the two relations and a similar cloud depth are also obtained in the Barnard 1 region in Perseus (Paper I).

The scatter in the $\langle \sigma \rangle_{\text{nt}}$ vs.\ size relation shows a trend with size. The smaller objects have larger scatter in $\langle \sigma \rangle _{\text{nt}}$ than bigger objects.
This scatter, which is several times the error in the means, could be due to a combination of true variation in the depth along the line of sight and of variations in the turbulent power.
The scatter with decreasing sizes rises because any line of sight is a specific realization of the turbulent power spectrum and the cloud depth;
hence statistical variation in the turbulent power is averaged out over larger objects (not for small objects).
Similarly, the calculation of $\Delta \text{V}_{\text{lsr}}$ averages over depth variations within larger objects.

\section{Filaments in Serpens Main}
\label{sect:filament}

\begin{figure}[t!]
\begin{center}
\includegraphics[scale=2.1]{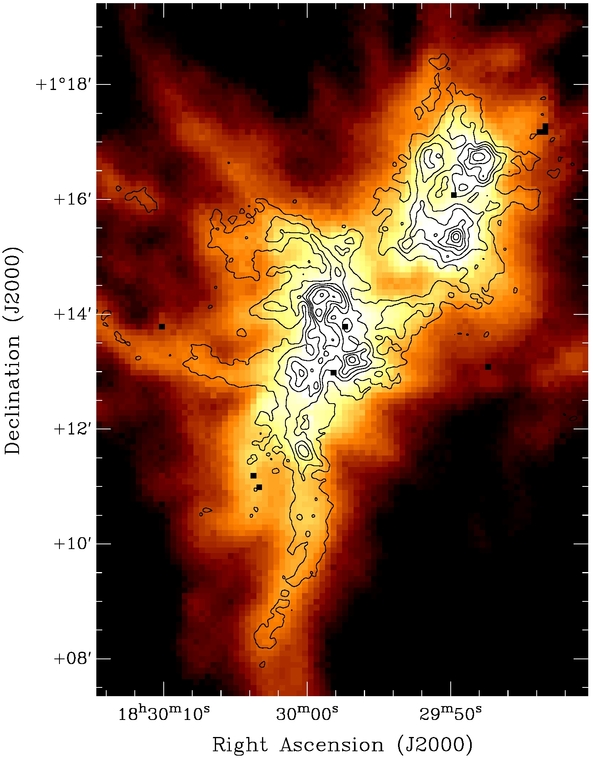}
\caption{\footnotesize
The integrated intensity map of \nh \ overlaid on a \textit{Herschel} 250 $\um$ image.
The \textit{Herschel} image is obtained from the \textit{Herschel} Science Archive (HSA).
The contour levels are at 20, 30, 40, 50, 60, 70, 80, 90\% of the peak value.
}
\label{fig:250_n2hp}
\end{center}
\end{figure}

As discussed in Sect.\ \ref{sect:mom0}, prominent filamentary structures are observed in Serpens Main, especially with the \nh \ observations. Figure~\ref{fig:250_n2hp} displays the \nh\ integrated intensity map overlaid on \textit{Herschel} 250~$\um$ emission. The \nh \ emission traces several of the most prominent filaments evident in the dust emission. Figure~\ref{fig:leaf} shows that the long southern filament is linked to several leaves and a branch; other filaments have single leaves or are only identified as branches because the emission is extended and low-intensity.

\begin{figure}[h]
\begin{center}
\includegraphics[scale=1.95]{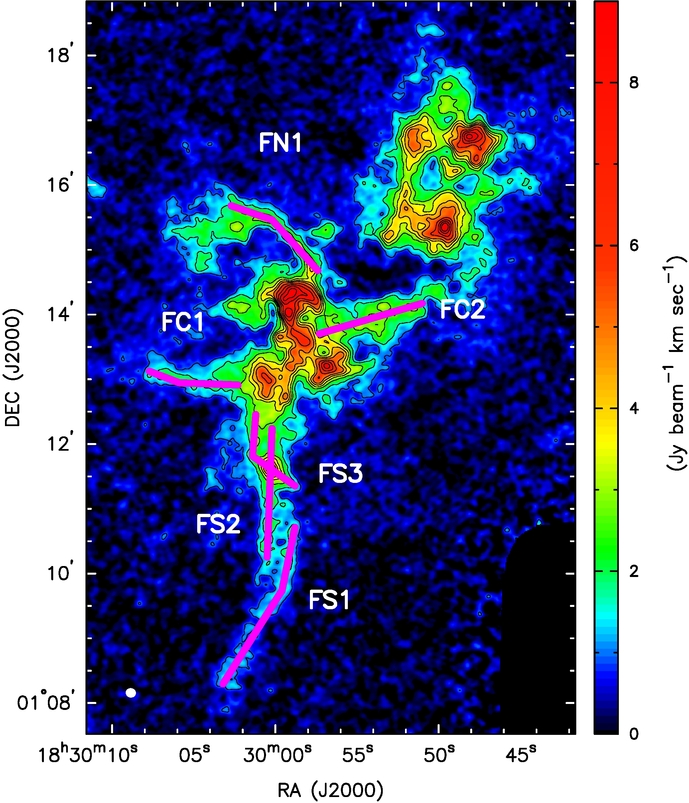}
\caption{\footnotesize
Six identified filaments on top of the \nh \ integrated emission (the color image; the same as Fig.~\ref{fig:serpens_n2hp}).
The contours are 4, 7, 11, 14, 17, 20, 25, 30, 35, 40$\times \sigma$ ($\sigma=0.24$ Jy~beam$^{-1}$~km~s$^{-1}$).
The synthesized beam ($7.7\arcsec \times 7.0\arcsec$) is drawn in the bottom-left corner.
}
\label{fig:fila_mom0}
\end{center}
\end{figure}

\begin{figure}
\begin{center}
\includegraphics[scale=3.0]{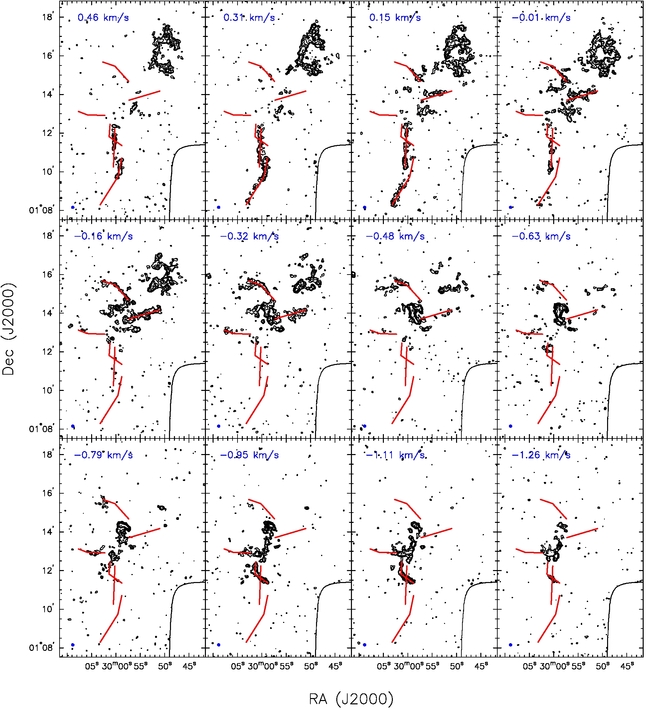}
\caption{\footnotesize
The identified filaments on channel maps of the isolated hyperfine component in the \nh \ emission.
Velocities associated with each channel are labeled in blue on the top of each channel.
The contours are $\pm3,\ 4.2,\ 6,\ 8.4,\ 12,\ 16.8,\ 24 \times \sigma$ ($\sigma=0.2$ Jy~beam$^{-1}$).
}
\label{fig:fila_chan}
\end{center}
\end{figure}

To better understand the physical properties of the filaments,
we identified six filaments as shown in Figure~\ref{fig:fila_mom0} by a thorough inspection of the morphologies in the integrated intensity map and the velocity channel maps of the isolated hyperfine component of \nh \ (Figure~\ref{fig:fila_chan}).
The filaments were identified based on (1) extended regions with signal-to-noise ratios higher than $3\sigma$ in both the channel maps and integrated intensity map, (2) emission in at least two adjacent channels, and (3) aspect ratios larger than 4.

\begin{deluxetable}{ccccccccccccccc}
\tablewidth{0pc}
\tablecolumns{15}
\tabletypesize{\tiny}
\setlength{\tabcolsep}{0.02in}
\tablecaption{Physical Properties of Filaments from the \nh \ emission}
\tablehead{
\colhead{Filament} & \colhead{Length} & \colhead{Width} & \colhead{Aspect}& \colhead{$\langle \text{V}_{\text{lsr}} \rangle$} & \colhead{$\nabla V$} & \colhead{$\langle \sigma \rangle$} & \colhead{$\langle \sigma \rangle_{\mathrm{nt}}$} & \colhead{$\langle \sigma \rangle_{\text{nt}}$/$c_s$} & \colhead{Subthermal Area} & \colhead{T} & \colhead{N$_{H_{2}}$} & \colhead{Mass} & \colhead{$M_{L}$} & \colhead{$M_{L,crit}$}  \\
\colhead{} & \colhead{(pc)} & \colhead{(pc)} & \colhead{ratio} & \colhead{(km~s$^{-1}$)} & \colhead{(km~s$^{-1}$~pc$^{-1}$)} & \colhead{(km~s$^{-1}$)} & \colhead{(km~s$^{-1}$)} & \colhead{} & \colhead{(\%)} & \colhead{(K)} & \colhead{($\times 10^{22}$ cm$^{-2}$)} & \colhead{(M$_{\sun}$)} & \colhead{(M$_{\sun}$~pc$^{-1}$)} & \colhead{(M$_{\sun}$~pc$^{-1}$)} \\
}
\startdata
FS1 & 0.33 & 0.04 & 8 & 8.27 & $0.8\pm 0.03$ & 0.13 & 0.12 & 0.54 & 95 & 11.5 & $4.3$ & 16.33 & 49.5 & 19.2  \\
FS2 & 0.24 & 0.03 & 8 & 8.14 & $1.5\pm 0.08$ & 0.24 & 0.23 & 1.04 & 60 & 12.1 & $7.9$ & 17.95 & 74.8 & 20.2  \\
FS3 & 0.17 & 0.03 & 6 & 7.05 & $0.8\pm 0.07$ & 0.23 & 0.22 & 1.02 & 70 & 12.3 & $9.3$ & 14.08 & 82.8  & 20.5  \\
FC1 & 0.17 & 0.04 & 4 & 7.35 & $4.8\pm 0.14$ & 0.25 & 0.24 & 1.11 & 57 & 12.0 & $3.5$ & 4.76 & 28.0 & 20.0  \\
FC2 & 0.20 & 0.05 & 4 & 7.82 & $0.7\pm 0.07$ & 0.21 & 0.20 & 0.92 & 67 & 13.4 & $5.2$ & 10.57 & 52.9 & 22.4  \\
FN1 & 0.21 & 0.03 & 7 & 7.79 & $3.2\pm 0.08$ & 0.27 & 0.21 & 1.18 & 42 & 14.2 & $3.6$ & 3.79 & 18.0 & 23.7  \\
\enddata
\label{tbl:t_filaments}
\end{deluxetable}

All the filaments identified in the \nh \ images are associated with the SE subcluster.
Three filaments (FS1, FS2, and FS3) are in the south and one filament (FN1) is in the north of the SE subcluster.
Two filaments (FC1 and FC2) are in the E-W direction.
FS2 and FS3 intersect in projection and a young stellar object, SMM11, is observed in the position where the two filaments cross (see Sect.\ \ref{sect:mom0}).
All six filaments have dust counterparts in the \textit{Herschel} 250~$\um$ emission;
however, FS1 and FS2 are seen as one filament in the \textit{Herschel} map.

We analyzed the morphological properties of these filaments based on the \nh \ emission: lengths, widths, and aspect ratios.
The results are summarized in Table~\ref{tbl:t_filaments}.
The filaments have lengths ranging from 0.17 to 0.33~pc (using distance $=415$ pc) with an average length of 0.22~pc.
The \nh \ lengths are significantly smaller than the parsec-long filaments found by \textit{Herschel} \citep{2010A&A...518L.102A}, but similar to several filaments in other star-forming sites revealed by molecular line observations \citep{2013ApJ...764L..26B,2013ApJ...766..115K,2013A&A...554A..55H}.  
We also estimated the averaged width of each filament by applying Gaussian fits to averaged intensity profiles along the direction perpendicular to the filament in the integrated intensity map.
On average, the filaments have widths of 0.036~pc (see Sect.~\ref{sect:width}).
The aspect ratios range from 4 to 7.

\begin{figure}[h]
\begin{center}
\includegraphics[scale=0.44]{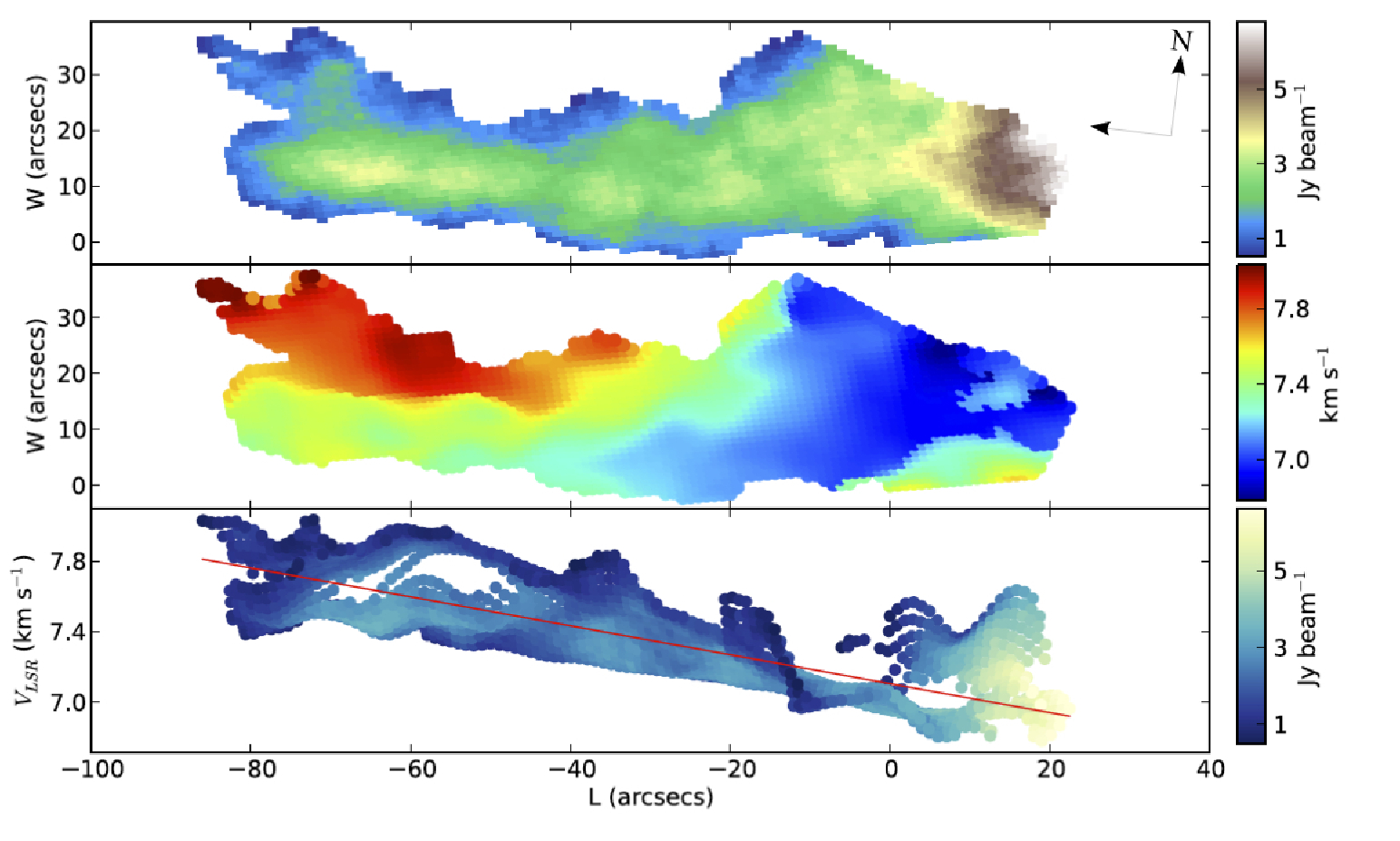}
\caption{\footnotesize
An example of FC1 demonstrating the fitting of velocity gradients.  
\textit{Top panel:} The \nh \ integrated intensity map of FC1. ``L" and ``W" show the spatial offsets along and perpendicular to the filament, respectively.  
\textit{Middle panel:} The centroid velocity map ($\text{V}_{\text{lsr}}$) of the filament. 
\textit{Bottom panel:} $\text{V}_{\text{lsr}}$ vs.\ L.  The velocity gradient is estimated based on a linear fit shown as the red, dashed line.
}
\label{fig:vgrad}
\end{center}
\end{figure}

We also analyzed six kinematic properties of these filaments:  averaged centroid velocity ($\langle \text{V}_{\text{lsr}} \rangle$), velocity variation ($\Delta \text{V}_{\text{lsr}}$), velocity gradient along the filament ($\nabla V$), averaged velocity dispersion ($\langle \sigma \rangle$), nonthermal velocity dispersion ($\langle \sigma \rangle_{nt}$), and $\langle \sigma \rangle_{nt}$ compared with the sound speed ($\langle \sigma \rangle_{nt}/C_{s}$).
The results are summarized in Table~\ref{tbl:t_filaments}. 
All these quantities were derived by analyzing each filament in the centroid velocity and velocity dispersion maps shown in Figure~\ref{fig:serpens_v}.
We removed the region with two velocity components in the intersection of FS2 and FS3 in the analysis (Fig.~\ref{fig:multiv}) to obtain more accurate estimates. 
Velocity gradients are estimated using linear fits along the filaments.
Figure~\ref{fig:vgrad} illustrates the fitting with an example of FC1.
All the filaments show velocity gradients along their major axes.
The magnitude of these gradients along the filaments range from 0.7 to 4.8 km~s$^{-1}$~pc$^{-1}$. 
For most of the filaments,
these velocity gradients are discontinued when approaching the SE subcluster.
FC1 and FN1 have the largest velocity gradients (4.8 and 3.2 km~s$^{-1}$), a factor of 3 to 5 larger than other filaments.

We estimated the nonthermal components of the velocity dispersions ($\langle \sigma \rangle_{\text{nt}}$). 
We used a gas temperature of 13 K for the filaments from the NH$_{3}$ observations \citep{2013A&A...553A..58L}.
A lower temperature for filaments compared with the overall cloud (assuming 20 K) is reasonable since the star formation activities are much less in the filaments than in the central regions of the subclusters.  
FN1 and FC1 have larger nonthermal velocity dispersions; the ratios $\langle \sigma \rangle_{nt}/c_{s}$ indicate that a large fraction of the gas in these two filaments is above the sonic level. 
On the other hand, FS1 is the most quiescent filament, with $\langle \sigma \rangle_{nt}/c_{s}$ about 0.5.
FC2, FS2, and FS3 are at the sonic level.

We also estimated H$_{2}$ column densities ($N_{H_{2}}$) and temperatures (T) of the filaments.
We performed a pixel-by-pixel SED fitting using the 160, 250, and 350 $\um$ data from \textit{Herschel}. For the fitting, the Herschel images were smoothed to a common resolution of 24 arcsec in the 350 $\um$ image using convolution kernels \citep{2008ApJ...682..336G,2011PASP..123.1218A}, and re-gridded to $10\arcsec$ pixels.
The filament regions were background subtracted using manually selected, nearby regions that represent local background emission.  
We assumed a grey-body emission with single temperature, and fit column densities and temperatures \citep[e.g.,][]{2010A&A...518L.106K,2010A&A...518L..92W}.
We also assumed a dust opacity of $\kappa_{\nu}= 0.1 \times (\nu / 1000 \ \text{GHz})^{\beta}$ cm$^{2}$~g$^{-1}$ \citep{1990AJ.....99..924B}. 
We assume $\beta=2$ \citep{1983QJRAS..24..267H} for filaments since filaments are less dense compared with the compact sources assuming $\beta=1.5$ (Sect.~\ref{sect:conti}).
The averaged temperatures and column densities on the filaments are summarized in Table~\ref{tbl:t_filaments}.

The filaments have dust temperatures ranging from 11.5 K to 14.2 K.
The averaged column densities range from $4\times 10^{22}$ to $9\times 10^{22}$ cm$^{-2}$, comparable to the column density estimates ($N_{H_{2}} \geq 10^{22}$ cm$^{-2}$) for forming prestellar/protostellar cores on the filaments in Aquila Rift \citep{2010A&A...518L.102A}.
We calculated the mass for each filament by summing the column densities from all the pixels.
Three filaments in the South (FS1, FS2, and FS3) have masses $\sim 15$ M$_{\sun}$ each,
and two filaments (FC1 and FN1) have lower masses $\sim 4$ M$_{\sun}$ each.

We also calculated ``mass per unit length" ($M_{L}$) along the filaments by dividing the masses and the lengths.
We further compared the mass per unit length ratios with critical values. 
Critical values are estimated by assuming isothermal, self-gravitating cylinders with no magnetic support: 
\begin{equation}
\label{equ:crit}
M_{L,crit} = 2c_{s}^{2}/G = 16.7~(\dfrac{T}{10K}) ~M_{\sun}~pc^{-1}
\end{equation}
\citep{1964ApJ...140.1056O}.
If a mass per unit length ratio is larger than $M_{L,crit}$, it becomes thermally supercritical and possibly undergoes gravitational contraction.
We used the averaged dust temperature on each filament derived from the SED fitting for each filament. 
FS1, FS2, and FS3 have their mass per unit length ratios significantly larger than the critical values ($M_{L} \sim 3.5\times M_{L,crit}$), 
while FC1 and FN1 have similar mass per unit length ratios with the critical values ($M_{L} \simeq M_{L,crit}$).
These results are summarized in Table~\ref{tbl:t_filaments}.

\section{Implications from Filaments}

\subsection{Filament Width}
\label{sect:width}

Recent observations from \textit{Herschel} suggested a characteristic FWHM width of $0.09\pm0.04$~pc derived from 278 filaments in eight regions (IC 5146, Orion B, Aquila Rift, Musca, Pipe Nebula, Polaris, Taurus and Ophiuchus; \citet{2013A&A...553A.119A}).
In Sect.~\ref{sect:filament}, we estimated the widths of filaments identified from our \nh \ data to be $0.036$~pc. This \nh \ width is about one-third of the characteristic width from \textit{Herschel}.
Indeed, the \textit{Herschel} filaments appear to be more extended in width compared with the \nh \ filaments in Serpens Main (Fig.\ \ref{fig:250_n2hp}).

\citet{manuel14} reported a similar width for a few filaments in Serpens South using \nh.
That study suggests that the narrower widths observed with \nh \ are possibly due to a combined effect of excitation conditions and chemical reactions.
At the centers of the filaments where densities are expected to be above 10$^{5}$ cm$^{-3}$, \nh \ emission depends almost linearly on its column density at a fixed temperature, and the \nh \ emission would be observed with sufficient excitation conditions in temperature and density for the transition.
Toward the edges where densities drop rapidly below 10$^{5}$ cm$^{-3}$, the emission efficiency could drop rapidly because of insufficient increase in the \nh \ column density to emit a fixed line temperature.  
Another possible mechanism for the narrower widths is from chemical effects \citep{2001ApJ...557..209B,2002ApJ...570L.101B}:
at regions less than CO depletion threshold ($2-6 \times 10^{4}$~cm$^{-3}$; \citet{2002ApJ...569..815T}), CO destroys \nh, resulting in a drop in the \nh \ abundance and hence a drop in the emission. 
However, the difference in the \nh \ widths and dust widths may as well be due to the \textit{Herschel} resolution not being sufficient to resolve 0.03~pc structures at the Serpens distance.  

While the discussion above corresponds to dust filaments composed of single \nh \ filaments, which most of the filaments in Serpens Main show, 
FS1 and FS2 are observed as two separate structures inside one \textit{Herschel} dust filament (Fig.~\ref{fig:250_n2hp}).
In this particular case, the width difference between the dust filament and the two molecular filaments is not likely due to excitation conditions nor chemical effects.
The \nh \ observations show substructures in the dust filament, resolving it into two quasi-parallel filaments with $\sim$0.03~pc width. 
This result, along with the obtained in the Serpens South filaments, suggests that filamentary structures found in some star-forming regions can be more complex than single filaments represented by \textit{Herschel}, and high angular resolution observations are needed to examine the widths of filaments in star-forming regions.

\subsection{Two Types of Filaments}

A number of recent studies have performed molecular line observations and revealed kinematic properties of filamentary structures toward nearby star-forming regions \citep{2010A&A...520A..49S,2013ApJ...766..115K,2013A&A...550A..38P,2013A&A...553A.119A} and IRDCs \citep{2012A&A...540A.104M,2014A&A...561A..83P}.
Filaments in some regions, including IRDC G14.225-0.506 and G035.39-00.33, show supersonic nonthermal velocity dispersions \citep{2013ApJ...764L..26B,2013MNRAS.428.3425H}, suggesting significant nonthermal contributions in the physical process of star formation.
On the other hand, subsonic to transonic nonthermal velocity dispersions with velocity coherence have been observed in regions such as L1517 \citep{2011A&A...533A..34H}, B5 \citep{2011ApJ...739L...2P}, and B213 in Taurus \citep{2013A&A...554A..55H}.
Most of our identified filaments (FC1 and FN1) in Serpens Main (Sect.~\ref{sect:filament}) show transonic velocity dispersions, while FS1 shows particularly subsonic velocity dispersions.

The properties of the filaments in the SE subcluster (Table~\ref{tbl:t_filaments}) show that there are two types of filaments in this region. 
Examples of one type are filaments FC1 and FN1 which show large velocity gradients along their major axis ($>3$ km~s$^{-1}$), small masses (4 M$_{\sun}$), and nearly critical mass per unit length ($M_{L} \simeq M_{L,crit}$; see Equ.~\ref{equ:crit}). 
In another type of filaments, we observe the opposite characteristics with small velocity gradients along the axis ($< 2$ km~s$^{-1}$), larger masses ($\sim$15 M$_{\sun}$), and supercritical mass per unit lengths ($M_{L} \simeq 3.5\times M_{L,crit}$), such as in filaments FS1, FS2, and FS3.
Filament FC2 appears to have intermediate characteristics although its properties are closer to those of the southern filaments. 

Each type of filaments appears to be spatially correlated:
FN1 and FC1 are northeast of the SE subcluster and the other filaments are south or west of it.
The correlations suggest that these filaments could be associated with larger structures.
The observed, dense filaments are in the intersection of large-scale structures possibly originated from large-scale turbulence \citep[e.g.,][]{2001ApJ...553..227P,2004RvMP...76..125M,2004ARA&A..42..211E,2007ARA&A..45..565M,2012A&ARv..20...55H}.

\begin{figure}[t!]
\begin{center}
\includegraphics[scale=2.1]{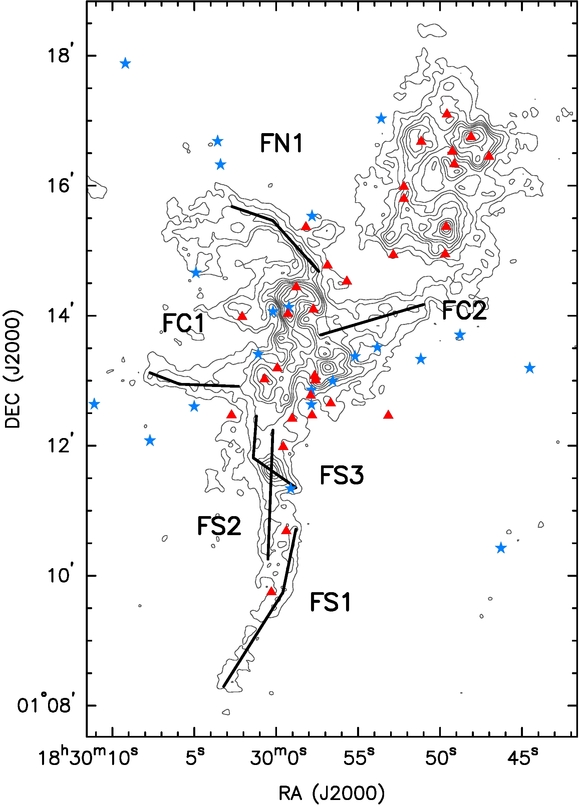}
\caption{
\footnotesize
A comparison between \nh \ filaments and YSOs.
Six identified filaments (Sect.~\ref{sect:filament}) are overlaid on the \nh \ integrated intensity map (contour levels: 5, 10, 15, 20, 25, 30, 40, 50, 60 $\times \sigma$; $\sigma=0.2$ km~s$^{-1}$).
YSOs identified by \citet{2009ApJS..181..321E} are shown:
red triangles are Class 0/I sources, and blue stars are Class II/III sources.
}
\label{fig:fila_yso}
\end{center}
\end{figure}

\subsection{The N$_{2}$H$^{+}$ Gas, Filaments, and YSOs}

Figure~\ref{fig:fila_yso} shows the comparison between the \nh \ gas and \textit{Spitzer} YSOs from \citet{2009ApJS..181..321E}.
Younger YSOs (Class 0 and I sources) are represented by red triangles, and more evolved YSOs (Class II and III sources) are represented by blue stars.
In general, younger YSOs are more closely related to the \nh \ emission and are concentrated in the two subclusters, while more evolved YSOs are distributed more widely.
This suggests that the \nh \ gas correlates with early stages of star formation.
Also, the NW subcluster contains primarily younger sources, and the SE subcluster is associated with both younger and more evolved YSOs, suggesting that the NW subcluster is younger than the SE subcluster.

Figure~\ref{fig:fila_yso} also shows a comparison between the \nh \ filaments and YSOs.
It is striking that a string of five YSOs is formed along FS1 and FS2,
while no YSOs are associated with FC1, FN1, and FC2.
FS1 and FS2 have supercritical mass per unit lengths ($M_{L}$ = 2.6 and 3.7 $\times$ $M_{L,crit}$, respectively),
while FC1 and FN1 have nearly critical mass per unit lengths.
These results suggest that stars are formed in situ along filaments that are gravitationally unstable.
Furthermore, the results could reflect that FS1 and FS2, which belong to the same type of filaments,
are more evolved than FN1 and FC1 in the other type.

\section{Summary}

We presented observations of \nh, \hcop, \hcn, and dust continuum at 3~mm~in Serpens Main from the CARMA Large Area Star Formation Survey (CLASSy).
The observations have an angular resolution of $\sim$ 7\arcsec \ and a spectral resolution of 0.16 km~s$^{-1}$. 
Our main conclusions are summarized below.

\begin{enumerate}
\item
\nh \ emission is concentrated in two subclusters, the NW subcluster and the SE subcluster.  
Prominent filamentary structures are observed in the SE subcluster, and the two southern filaments are resolved with high angular resolution for the first time.
\hcop \ and \hcn \ show more extended emission possibly from lower density gas.

\item
\nh \ line fitting shows that 
the NW subcluster has a fairly uniform velocity field, while the SE subcluster has more complicated velocity structures.
A majority of the regions (60\%) shows subsonic to sonic velocity dispersions in gas motions along the line of sight (assuming a gas temperature of 20 K).
The central region of the SE subcluster shows transonic to supersonic gas, while the surrounding filaments show more quiescent gas.  

\item
We identify 18 continuum sources at 3~mm.
The 3~mm sources are distributed in the NW and SE subclusters.
All eleven submillimeter cores from \citet{1999MNRAS.309..141D} have 3~mm counterparts.  

\item
We quantify the hierarchical, morphological, and kinematic properties of the \nh \ emission using a non-binary dendrogram analysis.
The dendrogram tree has 45 leaves and 19 branches.
Tree statistics (the mean branching level, mean path length, mean branching ratio) suggests that the SE subcluster has more complex hierarchical structure than the NW subcluster.
The complexity in the hierarchy is linked with star formation activity.

\item
The leaves and branches have a mean geometric size of 0.05 pc and 0.26 pc, respectively.  
Branches have larger $\text{V}_{\text{lsr}}$ variation ($\Delta \text{V}_{\text{lsr}}$) than the leaves, with a mean value of 0.29 km~s$^{-1}$ compared to 0.09 km~s$^{-1}$.
There is no obvious distinction in the velocity dispersions between the leaves and branches, and the mean value of all the objects is 0.2 km~s$^{-1}$, below the sonic level at 20 K.  

\item
The mean nonthermal velocity dispersion ($\langle \sigma \rangle_{\text{nt}}$) vs.\ size relation shows that $\langle \sigma \rangle_{\text{nt}}$ is nearly constant with size. 
The $\text{V}_{\text{lsr}}$ variation ($\Delta \text{V}_{\text{lsr}}$) vs.\ size relation shows a positive correlation. 
Assuming that an isotropic 3D turbulence dominates the gas motions,  
the mean nonthermal velocity dispersion reflects the nonthermal motions along the line of sight and the $\text{V}_{\text{lsr}}$ variation reflects the motions across different spatial scales on the plane of the sky;
the intersection of the two fitted power-law relations corresponds to an equal spatial scale in both dimensions.
It suggests that the cloud has a characteristic depth of 0.15~pc.   

\item 
We identify six filaments in the SE subcluster using the \nh \ data. 
The filaments typically have lengths of $\sim 0.2$ pc and \nh \ widths of $\sim 0.036$ pc.
This average width is about one third the characteristic width of 0.1~pc from \textit{Herschel} observations \citep[e.g.,][]{2013A&A...553A.119A}, in agreement with results for the \nh \ filaments in Serpens South by \citet{manuel14}.
The narrower widths in \nh \ may be due to a combination effect from excitation conditions and chemical reactions, 
or insufficient angular resolution from \textit{Herschel} to resolve substructures inside filaments.

\item
The \nh \ filaments can be divided into two types based on their properties. 
The first type shows large velocity gradients ($3-5$ km~s$^{-1}$~pc$^{-1}$), smaller masses ($\sim4$ M$_{\sun}$), and nearly critical mass-per-unit-length ratios.  
The second type shows the opposite properties with small velocity gradients ($0.8-1.5$ km~s$^{-1}$), larger masses ($\sim15$ M$_{\sun}$), and supercritical mass-per-unit-length ratios.
Each type of filament is spatially correlated, suggesting that the filaments in each type may be part of large-scale structures.

\item
Young YSOs (Class 0/I sources) identified with \textit{Spitzer} are more closely related to the \nh \ emission, while evolved YSOs (Class II and III sources) are distributed more widely.
This suggests that the \nh \ gas is linked with early stages of star formation.

\item
It is striking that a string of five YSOs is forming along FS1 and FS2, which have supercritical mass per unit lengths, while the filaments with nearly critical mass per unit lengths (FC1 and FN1) do not show associated YSOs.
This suggests that stars are formed in gravitationally unstable filaments.

\end{enumerate}

Overall, compared with the NW subcluster,
the SE subcluster shows more complicated velocity structures, a higher degree of hierarchy, and more prominent filamentary structures.
These results are consistent with previous studies that the SE subcluster is more evolved \citep[e.g.,][]{2007ApJ...669..493W}.

We thank the anonymous referee for valuable comments to improve the paper.  
CLASSy was supported by AST-1139950 (University of Illinois) and AST-1139998 (University of Maryland).
Support for CARMA construction was derived from the Gordon and Betty Moore Foundation, the Kenneth T. and Eileen L. Norris Foundation, the James S. McDonnell Foundation, the Associates of the California Institute of Technology, the University of Chicago, the states of Illinois, California, and Maryland, and the National Science Foundation.
Ongoing CARMA development and operations are supported by the National Science Foundation under a cooperative agreement, and by the CARMA consortium universities.

\clearpage

\bibliographystyle{apj}
\bibliography{paper}

\end{document}